\documentclass[draft,nofootinbib,pre,twocolumn,showpacs,showkeys,groupaddress,preprintnumbers,floatfix]{revtex4-1}

\usepackage{dynlearn}
\usepackage{physics}

\usepackage{mathrsfs}
\usepackage[english]{babel}
\usepackage{amssymb,amscd}

\newtheorem{proposition}{Proposition}

\newcommand{\e}{\mathrm{e}}

\DeclareMathOperator*{\argmax}{arg\,max}

\newcommand{\CP}[1]{\mathbb{C}P^{#1}}
\newcommand{\ID}{\mathfrak{D}}
\newcommand{\GE}{\mathfrak{h}_{\ID}}
\newcommand{\PHS}{\mathcal{P}(\mathcal{H}_S)}

\def\tbf #1 {\textbf{#1} }

\begin{document}

\def\ourTitle{%
Maximum Geometric Quantum Entropy
}

\def\ourAbstract{%
Any given density matrix can be represented as an infinite number of ensembles of pure states. 
This leads to the natural question of how to uniquely select one out of the many, apparently equally-suitable, possibilities. 
Following Jaynes’ information-theoretic perspective, this can be framed as an inference problem. 
We propose the Maximum Geometric Quantum Entropy Principle to exploit the notions of Quantum Information Dimension and Geometric Quantum Entropy. 
These allow us to quantify the entropy of fully arbitrary ensembles and select the one that maximizes it. 
After formulating the principle mathematically, we give the analytical solution to the maximization problem 
in a number of cases and discuss the physical mechanism behind the emergence of such maximum entropy ensembles.
%Density matrices capture all of a quantum system's statistics accessible
%through projective and positive operator-valued measurements. They do not
%specify how system statistics are created, however, as they neglect
%the physical realization of ensembles. Geometric quantum states---probability
%distributions of the system state conditioned on the environment state---were
%developed to track ensembles efficiently, using geometric quantum mechanics.
%Here, given knowledge of a density matrix, we show how to estimate the
%geometric quantum state using a maximum entropy principle based on a
%geometrically-appropriate quantum entropy. 
}

\def\ourKeywords{%
Quantum mechanics, geometric quantum mechanics, maximum entropy estimation,
density matrix
}

\hypersetup{
  pdfauthor={Fabio Anza},
  pdftitle={\ourTitle},
  pdfsubject={\ourAbstract},
  pdfkeywords={\ourKeywords},
  pdfproducer={},
  pdfcreator={}
}

%%%%%%%%%%%%%%%%%%%%%%%%%%%%%%%%%%%%%%%%%%%%%%%%%%%%%%%%%%%%%%%%%%%%%%%%%%%%%%%

\title{\ourTitle}

\author{Fabio Anza}
\email{fanza@ucdavis.edu}

\affiliation{Department of Mathematics Informatics and Geoscience, 
University of Trieste, Via Alfonso Valerio 2, Trieste, 34127, Italy}

\affiliation{Complexity Sciences Center and Physics Department,
University of California at Davis, One Shields Avenue, Davis, CA 95616}

\author{James P. Crutchfield}
\email{chaos@ucdavis.edu}

\affiliation{Complexity Sciences Center and Physics Department,
University of California at Davis, One Shields Avenue, Davis, CA 95616}

\date{\today}
\bibliographystyle{unsrt}

\begin{abstract}
\ourAbstract
\end{abstract}



\preprint{\arxiv{2008.08679 [quan-phys]}}

\date{\today}
\maketitle

% \tableofcontents

% \setstretch{1.1}

% Collides with table of contents formatting
% \listoffixmes

% {\bf Lead Paragraph:
% }

\section{Introduction}
\label{sec:Intro}

\paragraph*{Background.} Quantum mechanics defines a system's state $\ket{\psi}$ as an 
element of a Hilbert space $\mathcal{H}$. These are the \emph{pure states}. To account for
uncertainties in a system's actual state $\ket{\psi}$ one extends the
definition to \emph{density operators} $\rho$ that act on $\mathcal{H}$. These
operators are linear, positive semidefinite $\rho \geq 0$, self-adjoint $\rho =
\rho^\dagger$, and normalized $\Tr \rho = 1$. $\rho$ then is a pure state when
it is also a projector: $\rho^2 = \rho$.

The spectral theorem guarantees that one can always decompose a density
operator as $\rho = \sum_{i}\lambda_i \ket{\lambda_i}\bra{\lambda_i}$, where
$\lambda_i \in [0,1]$ are its eigenvalues and $\ket{\lambda_i}$ its
eigenvectors. \emph{Ensemble theory} \cite{Pathria2011,Greiner1995} gives the
decomposition's statistical meaning: $\lambda_i$ is the probability that the
system is in the pure state $\ket{\lambda_i}$. Together, they form $\rho$'s
eigenensemble $\mathcal{L}(\rho):=\left\{ \lambda_j,\ket{\lambda_j}\right\}_j$
which, putting degeneracies aside for a moment, is unique.
$\mathcal{L}(\rho)$, however, is not the only ensemble compatible with the
measurement statistics given by $\rho$. Indeed, there is an infinite number of
different ensembles that give the same density matrix: $\left\{
p_k,\ket{\psi_k}\right\}_k$ such that $\sum_k p_k \ket{\psi_k}\bra{\psi_k} =
\sum_j \lambda_j \ket{\lambda_j}\bra{\lambda_j}$. Throughout the following,
$\mathcal{E}(\rho)$ identifies the set of all ensembles of pure states
consistent with a given density matrix.

\paragraph*{Motivation.} Since the association $\rho \to \mathcal{E}(\rho)$ is
one-to-many, it is natural to ask whether a meaningful criterion to uniquely
select an element of $\mathcal{E}(\rho)$ exists. This is a typical inference
problem, and a principled answer is given by the maximum entropy principle
(MEP) \cite{Jaynes1957,Jaynes1957a,Cover2006}. Indeed, when addressing
inference given only partial knowledge, maximum entropy methods have
enjoyed marked empirical successes. They are broadly exploited in science
and engineering.

Following this lead, the following answers the question of uniquely selecting
an ensemble for a given density matrix by adapting the maximum entropy
principle. We also argue in favor of this choice by studying the dynamical 
emergence of these ensembles in a number of cases.

The development is organized as follows. Section \ref{sec:PreviousResults}
discusses the relevant literature on this problem. It also sets up language and
notation. Section \ref{sec:GQS} gives a brief summary of Geometric Quantum
Mechanics: a differential-geometric language to describe the states and
dynamics of quantum systems
\cite{Anza20a,Anza22,STROCCHI1966,Kibble1979,Heslot1985,Gibbons1992,Ashtekar1995,Ashtekar1999,Brody2001,Bengtsson2017,Carinena2007,Chruscinski2006,Marmo2010,Avron2020,Pastorello2015,Pastorello2015a,Pastorello2016,Clemente-Gallardo2013}.
Then, Section \ref{sec:max_geom_ent} introduces the technically pertinent
version of MEP---the Maximum Geometric Entropy Principle (MaxGEP). Section
\ref{sec:Discussion} discusses two mechanisms that can lead to the MaxGEP and
identifies different physical situations in which the ensemble can emerge.
Eventually, Section \ref{sec:Conclusion} summarizes what this accomplishes and
draws several forward-looking conclusions.

\section{Existing Results}
\label{sec:PreviousResults}

The properties and characteristics of pure-state ensembles is a vast and rich
research area, one whose results are useful across a large number of fields from
quantum information and quantum optics to quantum thermodynamics and quantum
computing, to mention only a few. This section discusses four sets of results
relevant to our purposes. This also allows introducing language and notation. 

First, recall Ref. \cite{HJW93} where Hughston, Josza, and Wootters gave a
constructive characterization of all possible ensembles behind a given density
matrix, assuming an ensemble with a finite number of elements. Second, Wiseman 
and Vaccaro in Ref. \cite{Wise01} then argued for a preferred ensemble via the 
dynamically-motivated criterion of a \emph{Physically Realizable} ensemble. Third, 
Goldstein, Lebowitz, Tumulka, and Zanghi singled out the Gaussian Adjusted 
Projected (GAP) measure as a preferred ensemble behind a density matrix in a 
thermodynamic and statistical mechanics setting \cite{Gold06}. Fourth, Brody and 
Hughston used one form of maximum entropy within geometric quantum 
mechanics \cite{Brody2000}. 

\paragraph*{HJW Theorem.} At the technical level, one of the most important
results for our purposes is the Hughston-Josza-Wootters (HJW) theorem, proved
in Ref. \cite{HJW93}, which we now summarize.

Consider a system with finite-dimensional Hilbert space $\mathcal{H}_S$
described by a density matrix $\rho$ with rank $r$: $\rho=
\sum_{j=1}^{r}\lambda_j \ket{\lambda_j}\bra{\lambda_j}$. We assume
$\mathrm{dim} \, \mathcal{H}_S := d_S = r$, since the case in which $d_S > r$
is easily handled by restricting $\mathcal{H}_S$ to the $r$-dimensional
subspace defined  by the image of $\rho$. Then, a generic ensemble $e_\rho \in
\mathcal{E}(\rho)$ with $d \geq d_S$ elements can be generated from
$\mathcal{L}(\rho)$ via linear remixing with a $d \times d_S$ matrix $M$ having
as columns $d_S$ orthonormal vectors. Then, $e_\rho = \left\{
p_k,\ket{\psi_k}\right\}$ is given by the following:
\begin{align*}
\sqrt{p_k}\ket{\psi_k} = \sum_{j=1}^{d_S}M_{kj}\sqrt{\lambda_j}\ket{\lambda_j}
  ~.
\end{align*}

Equivalently, one can generate ensembles applying a generic $d \times d$ unitary matrix 
$U$ to a list of $d$ non-normalized $d_S$-dimensional states in which the first 
$d_S$, $\left\{\sqrt{\lambda_j}\ket{\lambda_j}\right\}_{j=1}^{d_S}$, are proportional to the eigenvectors 
of $\rho$ while the remaining $d - d_S$ are simply null vectors: 
\begin{align*}
\sqrt{p_k}\ket{\psi_k}
  = \sum_{j=1}^{d_S} U_{kj}\sqrt{\lambda_j}\ket{\lambda_j}
  ~.
\end{align*}
Here, we must remember that $U$ is not an operator acting on $\mathcal{H}_S$,
but a unitary matrix mixing weighted eigenvectors into $d$ non-normalized
vectors.

The power of the HJW theorem is not only that it introduces a constructive way
to build $\mathcal{E}(\rho)$ ensembles, but that this way is complete. Namely,
all ensembles can be built in this way. This is a remarkable fact, which the
following sections rely heavily on.

\paragraph*{Physically Realizable Ensembles.} For our purposes, a particularly
relevant result is that of Wiseman and Vaccaro \cite{Wise01}. (See also
subsequent results by Wiseman and collaborators on the same topic
\cite{Kar11}.) The authors argue for a \emph{Physically Realizable} ensemble
that is implicitly selected by the fact that if a system is in a stationary
state $\rho_{ss}$, one would like to have an ensemble that is stable under the
action of the dynamics generated by monitoring the environment. This is clearly
desirable in experiments in which one monitors an environment to infer
properties about the system. While this is an interesting way to answer the
same question we tackle here, their answer is based on dynamics and limited to
stationary states. The approach we propose here is very different, being based
on an inference principle. This opens interesting questions related to
understanding the conditions under which the two approaches provide compatible
answers. Work in this direction is ongoing and it will be reported elsewhere.

\paragraph*{Gaussian Adjusted Projected Measure.} Reference \cite{Gold06} asks
a similar question to that here, but in a statistical mechanics and
thermodynamics context. Namely, viewing pure states as points on a
high-dimensional sphere $\psi \in \mathbb{S}^{2d_S-1}$, which probability
measure $\mu$ on $\mathbb{S}^{2d_S-1}$, interpreted as a \emph{smooth}
ensemble on $\mathbb{S}^{2d_S-1}$, leads to a thermal density matrix:
\begin{align*}
\rho_{th} = \int d\psi  \mu(\psi) \ket{\psi}\bra{\psi}~?
\end{align*}
Here, $\rho_{th}$ could be the microcanonical or the canonical density matrix.
Starting with Schr\"odinger's \cite{Schro27,Schrodinger1989} and Bloch's
\cite{Wal89} early work, the authors argue in favor of the Gaussian Adjusted
Projected (GAP) measure. This is essentially a Gaussian measure, adjusted and
projected to live on $\psi \in \mathbb{S}^{2d_S-1}$: 
\begin{align*}
GAP(\sigma) \propto e^{- \bra{\psi} \sigma^{-1} \ket{\psi}}
  ~.
\end{align*}

Written explicitly in terms of complex coordinates $\psi_j$, it is clear that this is a Gaussian measure with vanishing average $\mathbb{E}[\psi_j] = 0$ and 
covariance specified by $\mathbb{E}[\psi_j^{*} \psi_k] = \sigma_{jk}$. In
particular, $\sigma = \rho$ guarantees that $GAP(\rho)$ has $\rho$ as density
matrix. 

The GAP measure has some interesting properties \cite{Gold06,Gold16,Reim08} and, 
as we see in Section \ref{sec:max_geom_ent}, it is also closely related to one of our 
results in a particular case. Our results can therefore be understood as a generalization 
of the GAP measure. We will not delve deeper into this matter now, but comment on it later.

\paragraph*{Geometric Approach.} In 2000, Brody and Hughston performed the
first maximum entropy analysis for the ensemble behind the density matrix
\cite{Brody2000}, in a language and spirit that is quite close to those we use
here. Their result came before the definition of the GAP measure, but it is
essentially identical to it: $\mu(\psi) \propto \exp(-\sum_{j,k}
L_{jk}\psi_j^{*}\psi_k)$. Their perspective, however, is very different from
that in Ref. \cite{Gold06}, which is focused on thermal equilibrium
phenomenology. The work we do here, and our results, can also be understood as
a generalization of Ref. \cite{Brody2000}. Indeed, as we argued in Ref.
\cite{Anza22a} (and will show again in Section \ref{sec:max_geom_ent}) the
definition of entropy used (see Eq. (10) in Ref. \cite{Brody2000}) is
meaningful only in certain cases. In particular, when the ensemble has support
with dimension equal to the dimension of the state space of the system of
interest. In general, more care is required. 

\paragraph*{Summary.} We summarized four relevant sets of results on selecting
one ensemble among the infinitely many that are generally compatible with a
density matrix. Our work relies heavily on the HJW theorem \cite{HJW93}, and it
is quite different from the approach by Wiseman and Vaccaro \cite{Wise01}.
Moreover, it constitutes a strong generalization with respect to the results on
the GAP measure \cite{Gold06} in a thermal equilibrium context and with respect
to the analysis by Brody and Hughston in \cite{Brody2000}.

\section{Geometric Quantum States}
\label{sec:GQS}

Our maximum geometric entropy principle relies on a differential-geometric
approach to quantum mechanics called \emph{Geometric Quantum Mechanics} (GQM).
The following gives a quick summary of GQM and how its notion of Geometric
Quantum State \cite{Anza20a,Anza22,Anza22a} can be elegantly used to study
physical and information-theoretic aspects of ensembles. More complete
discussions are found in the relevant literature
\cite{STROCCHI1966,Kibble1979,Heslot1985,Gibbons1992,Ashtekar1995,Ashtekar1999,Brody2001,Bengtsson2017,Carinena2007,Chruscinski2006,Marmo2010,Avron2020,Pastorello2015,Pastorello2015a,Pastorello2016,Clemente-Gallardo2013}.

\paragraph*{Quantum State Space.} The state space of a finite-dimensional
quantum system with Hilbert space $\mathcal{H}_S$ is a projective Hilbert space
$\mathcal{P}(\mathcal{H}_S)$, which is isomorphic to a complex projective space
$\mathcal{P}(\mathcal{H}_S) \sim \mathbb{C}P^{d_S-1}:= \left\{ Z \in
\mathbb{C}^{d_S}: Z \sim \lambda Z, \lambda \in \mathbb{C}/{0}\right\}$.  Pure
states are thus in one-to-one correspondence with points $Z \in \CP{d_S-1}$.
Using a computational basis as reference basis
$\left\{\ket{j}\right\}_{j=1}^{d_S}$, $Z$ has homogeneous coordinates $Z =
(Z^1,\ldots,Z^{d_S})$ where $\ket{Z} = \sum_{j=1}^{d_S}Z^j \ket{j} \in
\mathcal{H}_S$. One of the advantages in using the geometric approach is that
one can exploit the symplectic character of the state space. Indeed, this
implies that the quantum state space $\CP{n}$ can essentially be considered as
a classical, although curved, phase space. With probability and phases being
canonically conjugated coordinates: $Z^j = \sqrt{p_j}e^{i\phi_j}$, we have
$\left\{ p_j,\phi_k\right\} = \delta_{jk}$. The intuition from classical
mechanics can then be used to understand the phenomenology of quantum systems.

\paragraph*{Observables.} Within GQM, observables are Hermitian functions from
$\CP{d_S-1}$ to the reals:
\begin{align*}
f_{\mathcal{O}}(Z):= \sum_{j,k=1}^{d_S} {Z^j}^{*}Z^k \mathcal{O}_{jk}/\sqrt{\sum_{h=1}^{d_S}\vert Z^h \vert^2}
  ~,
\end{align*}
where $\mathcal{O}_{jk} = \bra{j}\mathcal{O}\ket{k}$ are the matrix elements of
the Hilbert space self-adjoint operator $\mathcal{O}$. An analogous relation
holds for Positive Operator-Valued Measures (POVMs).

\paragraph*{Geometric Quantum States.} The quantum state space
$\mathcal{P}(\mathcal{H}_S)$ has a preferred metric $g_{FS}$ and a related
volume element $dV_{FS}$---the \emph{Fubini-Study volume element}. The details
surrounding these go beyond our present purposes. It is sufficient to give
$dV_{FS}$'s explicit form in the coordinate system we use for concrete
calculations. This is the  ``probability + phase'' coordinate system, given by $Z \leftrightarrow \left\{ (p_j,\phi_j)\right\}_{j=1}^{d_S}$: 
\begin{align*}
dV_{FS} & = \sqrt{\det g_{FS}} \prod_{j=1}^{d_S}
  dZ^j {dZ^j}^{*}  =  \prod_{j=1}^{d_S} \frac{dp_j d\phi_j}{2}
  ~.
\end{align*}

This volume element can be used to define integration. Indeed, calling
$\mathrm{Vol}[B]$ the volume of a set $B \subseteq \mathcal{P}(\mathcal{H}_S)$,
we have $\mathrm{Vol}[B]:=\int_B dV_{FS}$. In turn, this provides the
fundamental, unitarily invariant, notion of a uniform measure on the quantum
state space. This is the normalized Haar measure $\mu_{\mathrm{Haar}}$:
\begin{align*}
\mu_{\mathrm{Haar}}[B]
  = \mathrm{Vol}[B]/\mathrm{Vol}[\mathcal{P}(\mathcal{H}_S)]
  ~.
\end{align*}

$\mu_{\mathrm{Haar}}$ is a probability measure that weights all pure states
uniformly with the total Fubini-Study volume of the quantum state space.
Probability measures \cite{Pollard02,Kolmogorov1933} are the appropriate
mathematical formalization behind the physical notion of ensembles and they
formalize the concept of a Geometric Quantum State (GQS): A probability measure
on the (complex projective) quantum state space. For example, a pure state 
corresponds to a Dirac measure $\delta_{\psi}$ with support on a single point 
$\psi \in \mathcal{P}(\mathcal{H}_S)$, with Hilbert space representation $\ket{\psi}$.

\paragraph*{GQS as conditional probability measures.}

One way to embed the HJW theorem in this geometric context is the following. 

Any density matrix can be purified in an infinite number of different ways. A purification $\ket{\psi(\rho)}$ of $\rho$ is a pure state in a larger Hilbert space $\ket{\psi(\rho)} \in \mathcal{H}_S \otimes \mathcal{H}_E$ such that $\Tr_{E}\ket{\psi(\rho)}\bra{\psi(\rho)}$, where $\Tr_E$ is the partial trace over the additional Hilbert space $\mathcal{H}_E$. It is known that, for the purification to be achieved, $d_E \geq r$. Since we assume $r = d_S$, we have $d_E \geq d_S$. Any purification of $\rho$ will have a Schmidt decomposition of a specific type:
\begin{align*}
\ket{\psi(\rho)} = \sum_{j=1}^{d_S} \sqrt{\lambda_j} \ket{\lambda_j}\ket{SP_j}
  ~,
\end{align*}
where $\left\{\ket{SP_j} \in \mathcal{H}_E\right\}_{j=1}^{d_S}$ are the $d_S$
orthonormal ``Schmidt partners''. These can be extended to a full orthonormal basis 
on $\mathcal{H}_E$ by adding $d_E - d_S$ orthonormal vectors which are orthogonal 
to $\mathrm{span}(\left\{ \ket{SP_j}\right\}_j)$.

$\mathcal{L}(\rho)$ is therefore understood as the ensemble resulting from
conditioning on the Schmidt partners. Namely, when measuring the environment in
the basis $\left\{ \ket{SP_j}\right\}_{j=1}^{d_E}$ the state of the system after the measurement 
will be $\ket{\lambda_j}$ with probability $\lambda_j$. Its GQS is $\mu_{\mathcal{L}} = \sum_{j=1}^{d_S}\lambda_j \delta_{\lambda_j}$. 
If we now measure the environment in a generic basis, instead of using the Schmidt
partners, we generate a different ensemble. Calling
$\left\{\ket{v_\alpha}\right\}_{\alpha=1}^{d_E}$ one such basis, we have: 
\begin{align*}
\sum_{j=1}^{d_S} \sqrt{\lambda_j} \ket{\lambda_j}\ket{SP_j} = \sum_{\alpha=1}^{d_E} \sqrt{p_\alpha}\ket{\chi_\alpha} \ket{v_\alpha}
  ~,
\end{align*}
with $\sqrt{p_\alpha}\ket{\chi_\alpha} = \mathbb{I}_S \otimes \ket{v_\alpha}\bra{v_\alpha} \ket{\psi(\rho)}$, 
$\left\{ p_\alpha,\ket{\chi_\alpha}\right\}_{\alpha=1}^{d_E} \in
\mathcal{E}(\rho)$, and GQS $\mu = \sum_{\alpha=1}^{d_E} p_\alpha
\delta_{\chi_\alpha}$. 

Starting from the Schmidt partners, these bases are in one-to-one
correspondence with $d_E \times d_E$ unitary matrices acting on
$\mathcal{H}_E$: $\ket{v_\alpha} : = U\ket{SP_\alpha}$. And these, in turn, are
in one-to-one correspondence with the unitary matrices in the HJW theorem.
Therefore, they are an analogously complete classification of ensembles. The
reason for this slight rearrangement of things with respect to the HJW theorem
is that we now have an interpretation of $\ket{\chi_\alpha}$ as the
conditionally pure state of the system, conditioned on the fact that we make a
projective measurement $\left\{ \ket{v_\alpha}\right\}_{\alpha=1}^{d_E}$ on the
environment where the result $\alpha$ occurs with probability $p_\alpha$.

% If our system of interest interacts with an environment of 
%dimension $d_E$, and the two of them are in a pure state $\ket{\psi}$, then the GQS of our system \cite{Anza2022} 
%is determined by conditioning $\ket{\psi}$ on the result of a projective measurement on the environment $\left\{\ket{v_\alpha}\right\}_{\alpha=1}^{d_E}$. 
%If there are no correlations $\ket{\psi} = \ket{\psi_S}\ket{\psi_E}$ then the state of the system after we 
%perform a measurement on the environment is always the same, it is pure and, geometrically, it is represented
%as a point on $\mathcal{P}(\mathcal{H}_S)$. Therefore, as a measure, this is a Dirac measure $\mu = \delta_{\psi_S}$.

%In general, $\ket{\psi} = \sum_{\alpha=1}^{d_E}\sqrt{p_\alpha}\ket{\chi_\alpha}\ket{v_\alpha}$ will be 
%a correlated state. This leads to the ensemble $\left\{ p_\alpha,\ket{\chi_\alpha}\right\}_{\alpha=1}^{d_E}$,
%described by a GQS that is a convex sum of $d_E$ Dirac measures: $\mu = \sum_{\alpha=1}^{d_E}p_\alpha \delta_{\chi_\alpha}$.

\paragraph*{Quantifying quantum entropy.}
To develop entropy the following uses the setup in Refs.
\cite{Anza20a,Anza22,Anza22a} to study the physics of ensembles using geometric
quantum mechanics. Since the focus here is a maximum entropy approach to select
an ensemble behind a density matrix, it is important to have a proper
understanding of how to quantify the entropy of an ensemble or, equivalently,
of the GQS. 

First, we look at the statistical interpretation of the pure states which participate in  
the conditional ensembles $\left\{ p_\alpha,\ket{\chi_\alpha}\right\}$. The corresponding 
kets $\left\{ \ket{\chi_\alpha}\right\}$ are not necessarily orthogonal $\braket{\chi_\alpha}{\chi_\beta} \neq \delta_{\alpha \beta}$ 
so the states are not mutually exclusive, or distinguishable in the Fubini-Study sense. 
However, these states come with the classical labels $\alpha \to \ket{\chi_\alpha}$ associated 
to the outcomes of projective measurements on the environment. In this sense, if 
$\alpha \neq \beta$ we have a classical way to distinguish them and thus we can
understand how to interpret expressions like $-\sum_\alpha p_\alpha \log p_\alpha$.

Then, we highlight that the correct functional to use to evaluate the entropy of $\mu$ 
is not always the same. It depends on another feature of the ensemble, the
\emph{quantum information dimension}, which is conceptually related to the
dimension of its support in quantum state space. To illustrate the concept,
consider the following four GQSs of a qubit:
\begin{align*}
& \mu_1 = \delta_\psi && \mu_2 = \sum_{k} p_k \delta_{\psi_k} \\
& \mu_3 = \frac{1}{T}\int_0^T \!\! dt \delta_{\psi(t)} && \mu_4 = \mu_{\mathrm{Haar}}
  ~.
\end{align*}
Naturally, the entropy of $\mu_1$ vanishes, since there is no uncertainty. The
system inhabits only one pure state, $\psi$. The entropy of $\mu_2$ is already
nontrivial to evaluate. Indeed, while one obvious way is to use the functional
$-\sum_k p_k \log p_k$, it is also very clear that this notion of entropy does
not take into account the location of the points $\psi_k \in
\mathcal{P}(\mathcal{H}_S)$. Intuitively, if all these points are close to each
other, we would like our entropy to be smaller than in the case in which all
the points are uniformly distributed on $\psi_k \in \mathcal{P}(\mathcal{H}_S)$.

The entropy of $\mu_3$ is perhaps the most peculiar, but it illustrates the
points in the best way. Let's assume that our qubit is evolving with a
Hamiltonian $H$ such that $E_1 - E_0 = \hbar \omega$. Then, $\psi(t) =
(\sqrt{1-p_0},\sqrt{p_0} e^{i\phi_0 - \omega t})$. If we aggregate the
time average and look at the statistics we obtain, it is clear that the
variable $p$ is a conserved quantity---$p(t) = p_0$. While $\phi(t) = \phi_0 -
\omega t$ is an angular variable that, over a long time, will be uniformly
distributed in $[0,2\pi]$. This means $\lim_{T \to \infty}\mu_3 =
\frac{1}{2\pi}\delta^{p}_{p_0}$, where $\delta^p_{p_0}$ is a Dirac measure over
the first variable $p$ with support on $p=p_0$. How do we evaluate the
entropy of $\mu_3$?

While to evaluate $\mu_4 =\mu_{\mathrm{Haar}}$ we simply integrate over the
whole state space, and obtain $\log \mathrm{Vol}(\mathbb{C}P^1)$, this does not
work for $\mu_3$. Indeed, with respect to the full, 2D quantum state space
$(p,\phi) \in [0,1] \times [0,2\pi]$, the distribution clearly lives on a $1D$
line, which is a measure-zero subset.

To properly address all these different cases a more general approach is
needed. Reference \cite{Anza22a} adapted previous work by Renyi to probability
measures on a quantum state space. This led to the notions of \emph{Quantum
Information Dimension} $\mathfrak{D}$ and \emph{Geometric Quantum Entropy}
$H_{\mathfrak{D}}$ that address these issues and properly evaluate the entropy
in all these cases. We now give a quick summary of the results in Ref.
\cite{Anza22a}.

\paragraph*{Quantum Information Dimension and Geometric Entropy.}
Thanks to the symplectic nature of $\mathcal{P}(\mathcal{H}_S)$, the quantum
state space is essentially a curved, compact, classical phase space. We can therefore
apply classical statistical mechanics to it, using
$\left\{(p_j,\phi_j)\right\}_{j=1}^{d_S}$ as canonical coordinates. Since, the
Fubini-Study volume is $dV_{FS} \propto \prod_j dp_j d\phi_j$, we can
coarse-grain $\mathcal{P}(\mathcal{H}_S)$ by partitioning it into phase-space
cells $\mathcal{C}_{\vec{a},\vec{b}}$:
\begin{align*}
\mathcal{C}_{\vec{a}\vec{b}} = \prod_{j=1}^{d_S-1} \left[\frac{a_j}{N},\frac{a_{j}+1}{N}\right] \times 2\pi \left[\frac{b_j}{N},\frac{b_j+1}{N} \right]
\end{align*}
of equal Fubini-Study volume $\mathrm{Vol}\left[
\mathcal{C}_{\vec{a}\vec{b}}\right] =
\frac{\mathrm{Vol[\mathcal{P}(\mathcal{H}_S)]}}{N^{2({d_S-1})}} =
\epsilon^{-2(d_S-1)}$, where $\vec{a} = (a_1,\ldots,a_{d_S-1}),\vec{b} =
(b_1,\ldots,b_{d_S-1})$ and $a_j,b_j = 0,1,\ldots,N$.

The coarse-graining procedure produces a discrete probability distribution
$q_{\vec{a}\vec{b}} := \mu[\mathcal{C}_{\vec{a}\vec{b}}]$, for which we can
compute the Shannon entropy: 
\begin{align*}
H[\epsilon]:= - \sum_{\vec{a},\vec{b}} q_{\vec{a}\vec{b}}\log q_{\vec{a}\vec{b}}
  ~.
\end{align*}
As we change $\epsilon = 1/N \to 0$, the degree of coarse-graining changes accordingly. The scaling behavior of $H[\epsilon]$ provides structural
information about the underlying ensemble. Indeed, since one can prove that for $\epsilon \to 0$, $H[\epsilon]$ has asymptotics
\begin{align*}
H[\epsilon] \sim_{\epsilon \to 0} \mathfrak{D} \left( - \log \epsilon\right) + \mathfrak{h}_{\mathfrak{D}}
  ~,
\end{align*}
two quantities define its scaling behavior: $\mathfrak{D}$ is the quantum information dimension and 
$\mathfrak{h}_{\mathfrak{D}}$ is the geometric quantum entropy. Their explicit definitions are:
\begin{subequations}
\begin{align}
& \mathfrak{D}:= \lim_{\epsilon \to 0} \frac{H[\epsilon]}{- \log \epsilon}~, \\
& \mathfrak{h}_{\mathfrak{D}}:= \lim_{\epsilon \to 0} \left( H[\epsilon] + \mathfrak{D} \log{\epsilon} \right)
  ~.
\end{align}
\end{subequations}

Note how this keeps the dependence of the entropy on the information dimension
explicit. This clarifies how, only in certain cases, one can use the continuous
counterpart of Shannon's discrete entropy. In general, its exact form depends
on the value of $\mathfrak{D}$ and it cannot be written as an integral on the
full quantum state space with the Fubini-Study volume form.

\section{Principle of Maximum Geometric Quantum Entropy}
\label{sec:max_geom_ent}

This section presents a fine-grained characterization of selecting an ensemble
behind a given density matrix.  This leverages both the HJW theorem and
previous results by the authors. First, we note that $\ID$ foliates
$\mathcal{E}(\rho)$ into non-overlapping subsets $\mathcal{E}_{\ID}(\rho)$
collecting all ensembles $\mu$ at given density matrix $\rho$ and with
information dimension $\ID$:
\begin{align*}
& \mathcal{E}_{\ID}(\rho) \cap \mathcal{E}_{\ID^{'}}(\rho) = \delta_{\ID,\ID^{'}} \mathcal{E}_{\ID}(\rho)~, && \mathcal{E}(\rho) = \cup_{\ID} \mathcal{E}_{\ID}(\rho)
  ~.
\end{align*}
As argued above, ensembles with different $\ID$ pertain to different physical
situations. These can be wildly different. Therefore, we often want to first select 
the $\ID$ of the ensemble we will end up with and then choose that with the 
maximum geometric entropy. Thus, here we introduce the principle of maximum 
geometric entropy at fixed information dimension.

\begin{proposition}[Maximum Geometric Entropy Principle]
Given a system with density matrix $\rho$, the ensemble $\mu_{ME}^{\ID}$ that
makes the fewest assumptions possible about our knowledge of the ensemble among
all elements of $\mathcal{E}(\rho)$ with fixed information dimension dimension
$\ID$ is given by:
\begin{align*}
\mu_{ME}^{\ID}:= \argmax_{\mu \in \mathcal{E}_{\ID}(\rho)} \GE
  ~.
\end{align*}
\end{proposition}

Several general comments are in order. First, we note that $\mu_{ME}^{\ID}$
might not be unique. This should not come as a surprise. For example, with
degeneracies, even the eigenensemble is not unique. Second, the optimization
problem defined above is clearly constrained: the resulting ensemble has to be
normalized and the average of ${(Z^{j})}^{*}Z^{k}$ must be $\rho_{jk}$. Calling
$\mathbb{E}_\mu[A]$ the state space average of a function $A$ done with the 
GQS $\mu$, these two constraints can be written as $\mathfrak{C}_{1} := \mathbb{E}_{\mu}[1] - 1 = 0$
and $\mathfrak{C}^{\rho}_{jk}:=\mathbb{E}_{\mu}[{(Z^{j})}^{*}Z^{k}] - \rho_{jk} = 0$.
Using Lagrange multipliers, we optimize
$\Lambda[\mu,\gamma_1,\left\{\gamma_{jk} \right\}]$ defined as:
\begin{align*}
\Lambda[\mu,\gamma_1,\left\{\gamma_{jk} \right\}]:=\GE\left[\mu\right] + \gamma_1 \mathfrak{C}_1 + \sum_{j,k}\gamma_{jk}\mathfrak{C}^{\rho}_{jk} 
  ~.
\end{align*}
While the vanishing of $\Lambda$'s derivatives with respect to the Lagrange
multipliers $\gamma_1,\gamma_{jk}$ enforces the constraints
$\mathfrak{C}_{1}=\mathfrak{C}_{jk}^{\rho}=0$, derivatives with respect to
$\mu$ give the equation whose solution is the desired ensemble $\mu_{ME}^{\ID}$.
We also note that the $\left\{ \gamma_{jk}\right\}$ are not all independent.
This is due to the fact that $\rho$ is not an arbitrary matrix: $\Tr \rho = 1$,
$\rho \geq 0$, and $\rho^{\dagger}=\rho$. A similar relation holds for $\gamma_{jk}$.

To illustrate its use, we now solve this optimization problem in a number of
relevant cases. In discussing them, it is worth introducing additional
notation. Since we often use canonically conjugated coordinates, $\left\{
(p_j,\phi_j)\right\}_{j=1}^{d_S}$, we introduce vector notation
$(\vec{p},\vec{\phi})$, with $\vec{p} \in \Delta_{d_S-1}$ and $\vec{\phi} \in
\mathbb{T}^{d_S-1}$, where $\Delta_{d_S-1}$ is the $(d_S-1)$-dimensional probability
simplex and $\mathbb{T}_{d_S-1}$ is the $(d_S-1)$-dimensional torus. Analogously, we
introduce the Dirac measures $\delta^{\vec{p}}_{\vec{x}}$ and
$\delta^{\vec{\phi}}_{\vec{\varphi}}$ with support on $\vec{x} \in
\Delta_{d_S}$ and $\vec{\varphi} \in \mathbb{T}_{d_S}$, respectively.

\subsection*{Finite Environments: $\ID = 0$}

If $\ID = 0$ then the support of the ensemble is made by a number of points
which is a natural number. That is, there exists $N \in \mathbb{N}$ such that
$\mu^{\ID = 0}_{ME} = \sum_{\alpha = 1}^N p_\alpha \delta_{\chi_\alpha}$, with
$\mathfrak{h}_0 = - \sum_{\alpha=1}^N p_\alpha \log p_\alpha$. Note how this is
the HJW theorem's domain of applicability. And, this allows us to give a
constructive solution.

We start by noting that $N$ also foliates $\mathcal{E}_{\ID = 0}$ into
non-overlapping sets in which the ensemble consists of \emph{exactly} $N$
elements. We call this set $\mathcal{E}_{0,N}(\rho)$ and it is such that
$\mathcal{E}_{0,N}(\rho) \cap \mathcal{E}_{0,N^{'}}(\rho) =
\delta_{N,N^{'}}\mathcal{E}_{0,N}(\rho)$, with $\mathcal{E}_{\ID=0}(\rho) =
\cup_{N \geq d_S} \mathcal{E}_{0,N}(\rho)$. Within $\mathcal{E}_{0,N}(\rho)$,
we can use the HJW theorem with the interpretation in which the ensemble is the
conditional ensemble. Here, $p_\alpha$ and $\chi_\alpha$ are generated by
creating a purification of dimension $N$, in which the first $d_S$ elements of
the basis $\left\{\ket{SP_j}\right\}_{j=1}^{d_S}$ are fixed and the remaining
$N-d_S$ are free. We denote the entire basis of this type with the same symbol
but a different label: $\left\{\ket{SP_\alpha}\right\}_{\alpha=1}^{N}$. The
ensemble we get if we measure it is the eigenensemble $\mathcal{L}(\rho)$.

However, measuring in a different basis yields a general ensemble, with probabilities
$p_\alpha = \bra{\psi(\rho)} \mathbb{I}_S \otimes \ket{v_\alpha}\bra{v_\alpha}\ket{\psi(\rho)} = \sum_{j=1}^{d_S} \lambda_j \left\vert \braket{SP_j}{v_\alpha}\right\vert^2$
and states $\ket{\chi_\alpha} =  \sum_{j=1}^{d_S} \sqrt{\lambda_j} \frac{\braket{v_\alpha}{SP_{j}}}{\sqrt{p_\alpha}} \ket{\lambda_j}$.
With $\mathfrak{h}_0 = - \sum_\alpha p_\alpha \log p_\alpha$ the absolute maximum is attained at $p_\alpha = 1/N$. We now show,
constructively, that this is always achievable while still satisfying the constraints $\mathfrak{C}_1 = \mathfrak{C}^{\rho}_{ij} = 0$, thus
solving the maximization problem.

This is achieved by measuring the environment in a basis that is unbiased with respect to the 
Schmidt partner basis: 
\begin{equation}
\ket{v_\alpha}: \braket{v_\alpha}{SP_\beta} = \frac{e^{i\theta_{\alpha \beta}}}{\sqrt{N}}~\forall \alpha,\beta=1,\ldots,N.\label{eq:MUB_basis}
\end{equation}
One such basis can always be built starting from $\left\{\ket{SP_\alpha}\right\}_\alpha$ by exploiting the properties of the
Weyl-Heisenberg matrices via the clock-and-shift construction \cite{Vourdas04}. This is true for all $N \in \mathbb{N}$. When 
$N = \prod_{k}n_k^{N_k}$ with $n_k$ primes and $N_k$ some integers, the finite-field algorithm \cite{Bengtsson07,Lawrence02} can be 
used to build a whole suite of $N$ bases that are unbiased with respect to the Schmidt partner basis. 
This leads to $\ket{\chi_\alpha} = \sum_{j=1}^{d_S} \sqrt{\lambda_j} e^{i\theta_{\alpha j}}\ket{\lambda_j}$ and to:
\begin{align*}
&\mu_{ME}^{\ID=0} = \delta^{\vec{p}}_{\vec{\lambda}} \frac{1}{N}\sum_{\alpha=1}^{N} \delta^{\vec{\phi}}_{\vec{\theta}_\alpha}~, &&\mathfrak{h}_0 = \log N
  ~,
\end{align*}
with $\vec{\lambda} = (\lambda_1,\ldots,\lambda_{d_S})$ and $\vec{\theta}_\alpha = (\theta_{\alpha0},\ldots,\theta_{\alpha d_S})$.

To conclude this subsection we simply have to show that this ensemble satisfies the constraints: 
$\mathfrak{C}_1=0$, and that the density matrix given by $\mu_{ME}^{\ID=0}$ is $\rho$, giving $\mathfrak{C}_{jk}^{\rho}=0$:
\begin{align*}
\left[\sigma_{ME}\right]_{jk}&:=\mathbb{E}_{\mu_{ME}^{\ID=0}}\left[ \left(Z^{j}\right)^{*}Z^{k}\right] = \frac{\sqrt{\lambda_j \lambda_k}}{N}\sum_{\alpha=1}^{N} e^{i(\theta_{\alpha j}- \theta_{\alpha k})} \nonumber\\
& = \sqrt{\lambda_j \lambda_k} \delta_{jk} = \lambda_k \delta_{jk} = \rho_{jk}
  ~.
\end{align*}
Here, the key property used is that $\frac{1}{N}\sum_{\alpha=1}^N
e^{i(\theta_{\alpha \gamma} - \theta_{\alpha \beta})} = \bra{SP_\beta}
\sum_{\alpha =1}^N  \ketbra{v_\alpha}{v_\alpha} \ket{SP_{\gamma}}=\delta_{\beta
\gamma}$, which comes from Eq. (\ref{eq:MUB_basis}) and the fact that $\left\{ \ket{v_\alpha}\right\}_{\alpha}$
is a basis.

\subsection*{Full support: $\ID = 2(d_S-1)$}

The second case of interest is the one in which the quantum information dimension takes 
the maximum value possible, namely $\ID = 2(d_S-1)$. Then, the GQS's support has
the same dimension as the full quantum state space and the optimization problem
is also tractable. This is indeed the case solved by Brody and Hughston
\cite{Brody2000}. We do not reproduce the treatment here, which is almost
identical in the language of GQM. Rather, we discuss some of its physical
aspects from the perspective of conditional ensembles.

If $\ID = 2(d_S - 1)$ and there are no other constraints aside from
$\mathfrak{C}_1$ and $\mathfrak{C}_{jk}$, the measure $\mu_{ME}^{2(d_S-1)}$ can
be expressed as an integral with a density $q_{ME}$ with respect to the
uniform, normalized, Fubini-Study measure $dV_{FS}$:
\begin{align*}
\mu_{ME}^{2(d_S-1)}[A] = \int_A \!\! dV_{FS}~q_{ME}(Z)
  ~.
\end{align*}
And, its geometric entropy $\mathfrak{h}_{2(d_S-1)}$ is the continuous counterpart of Shannon's functional, on the quantum state space:
\begin{align*}
\mathfrak{h}_{2(d_S-1)} = -\int_{\mathcal{P}(\mathcal{H}_S)}\!\!\!\!\!\! dV_{FS}~q(Z) \log q(Z)
  ~.
\end{align*}
This was proven in Ref. \cite{Anza22a}. Hereafter, with a slight abuse of
language, we refer to both $\mu_{ME}^{2(d_S-1)}$ and the density $q_{ME}(Z)$ as
an ensemble or the GQS.

The maximization problem leads to:
\begin{align*}
& q_{ME}(Z) = \frac{1}{Q_{2(d_S-1)}(\rho)}e^{-\sum_{jk}\gamma_{jk} (Z^{j})^{*}Z^k}~,\label{eq:me_max_D}\\
& Q_{2(d_S-1)}(\rho) \coloneqq \int_{\mathcal{P}(\mathcal{H}_S)}\!\!\!\!\!\! dV_{FS} \,\,  e^{-\sum_{jk}\gamma_{jk} (Z^{j})^{*}Z^k}
\end{align*}
and Lagrange multipliers $\left\{ \gamma_{jk}\right\}$ are the solution of the nonlinear equations $-\frac{\partial \log Q}{\partial \gamma_{jk}} = \rho_{jk}$.
We note how, using as reference basis the eigenbasis $\left\{\ket{\lambda_j}\right\}_{j=1}^{d_S}$ of $\rho$ and $Z \leftrightarrow (\vec{p},\vec{\phi})$ as coordinate system
reveals that $\frac{\partial \log Q}{\partial \gamma_{jk}} = 0$ when $j\neq k$ and $-\frac{\partial \log Q}{\partial \gamma_{jj}}=\lambda_j$. Thus, in this coordinate system
the dependence of $\mu_{ME}^{2(d_S-1)}$ on the off-diagonal Lagrange multipliers disappears and we retain only the diagonal ones $\gamma_{jj}$. 

Moving to a single label $\gamma_{jj} \to \gamma_{j}$ and using a vector
notation:
\begin{align*}
& q^{2(d_S-1)}_{ME}(\vec{p},\vec{\phi}) = \frac{1}{Q_{2(d_S-1)}(\vec{\tau})}e^{\vec{\tau} \cdot \vec{p}}~,\\
& Q_{2(d_S-1)}(\vec{\tau}) \coloneqq \int_{\Delta_{d_S-1}}\!\!\!\!\!\! d\vec{p} \,\,  e^{\vec{\tau} \cdot \vec{p}}\\
& \vec{\tau} \coloneqq \left( \gamma_{d_S}-\gamma_1,\gamma_{d_S}-\gamma_2,\ldots,\gamma_{d_S}-\gamma_{d_S-1}\right)
\end{align*}
Here, $Q(\vec{\tau})$ is the normalization function (a partition function). Its
exact expression can be derived analytically and it is given in Appendix \ref
{App:DetailedCalculation}.

We can see how $\mu_{ME}^{2(d_S-1)}$ is the product of an exponential measure
on the probability simplex $\Delta_{d_S-1}$ and the uniform measure on the
high-dimensional torus of the phases $\mathbb{T}_{d_S-1}$. This leads to the
following geometric entropy $\mathfrak{h}_{2(d_S-1)}$:
\begin{align}
&\mathfrak{h}_{2(d_S-1)}(\vec{\tau}) = \log Q_{2(d_S-1)}(\vec{\tau}) - \vec{\tau} \cdot \vec{\lambda}
  ~.
\end{align}
In this case the explicit expression of the Lagrange multipliers $\vec{\tau}$
satisfying the constraints, which was previously unknown, can be found
analytically. This is reported in Appendix \ref{App:Tau}.

We note that this exponential distribution on the probability simplex was
recently proposed within the context of statistics and data analysis in
Ref. \cite{GordRodr20}. Moreover, the exponential form associated to the maximum
entropy principle is reminiscent of thermal behavior. Indeed, the shape of this
distribution is closely related to the geometric canonical ensemble, see
Refs. \cite{Brody2000,Brody2001,Anza22}. However, the value of the Lagrange
multipliers is set by a different constraint, in which we fix the average
energy rather than the whole density matrix.

\subsection*{Integer, but otherwise arbitrary, $\ID$}

While one expects $\ID$ to be an integer, there are GQSs that have fractal
support, thus exhibiting a noninteger $\ID$. This was shown in
Ref. \cite{Anza22a}. This section discusses the generic case in which
$\ID \neq 0,2(d_S-1)$, but it is still an integer. Within
$\mathcal{E}_{\ID}(\rho)$, our ensemble $\mu_{ME}^{\ID}$ has support on a
$\ID$-dimensional submanifold of the full $2(d_S-1)$-dimensional quantum state
space, where it has a density.  Reference \cite{Anza22a} discusses, in detail,
the case in which $\ID=1$ and $d_S = 2$. Here, we generalize the procedure to
arbitrary $\ID$ and $d_S$.

If the support of $\mu_{ME}^{\ID}$ is contained in a submanifold of dimension
$\ID < 2(d_S-1)$, which we call $\mathcal{S}_{\ID}$, we can project the
Fubini-Study metric $g^{FS}$ down to $\mathcal{P}(\mathcal{H}_S)$ to get
$g_{\mathcal{S}}^{FS}$. Let's call $X^{j}: \xi^a \in \mathcal{S}_{\ID} \to
X^j(\xi^a) \in \mathcal{P}(\mathcal{H}_S)$ the functions which embed
$\mathcal{S}_{\ID}$ into the full quantum state space $\PHS$. Then the metric
induced on $\mathcal{S}_{\ID}$ is $g^{\mathcal{S}}_{ab} =\sum_{j,k}\partial_a
X^{j} \partial_b X^{k} g_{jk}^{FS}$, where $\partial_a := \partial/\partial \xi^a$. Note that here we are using the ``real
index'' notation even for coordinates $X^j$ on $\PHS$. While $\PHS$ is a
complex manifold, admitting complex homogeneous coordinates, we can always use
real coordinates on it. Then $g^{\mathcal{S}}$ induces a volume form
$d\omega^{\xi}_{\mathcal{S}} = \omega_{\mathcal{S}}(d\xi) = \sqrt{\det
g^{\mathcal{S}}}d\xi$, where $d\xi$ is the Lebesgue measure on the
$\mathbb{R}^{\ID}$ which coordinatizes $\mathcal{S}_{\ID}$.  Then,
$\mu_{ME}^{\ID}$ can be written as:
\begin{align*}
\mu_{ME}^{\ID}[A \in \mathcal{S}_{\ID}] = \int_{A} \!\!\! d\omega^{\xi}_{\mathcal{S}} f(\xi)
  ~.
\end{align*}

Eventually, this leads to:
\begin{align*}
\mathfrak{h}_{\ID} = - \int_{\mathcal{S}_\ID}\!\!\! d\omega^{\xi}_{\mathcal{S}}~f(\xi) \log f(\xi)
  ~.
\end{align*}
This allows rewriting the constraints explicitly in a form that involves only probability densities on $\mathcal{S}_{\ID}$:
\begin{subequations}
\begin{align*}
\mathfrak{C}_1 &= \mu_{ME}^{\ID}[\mathcal{S}_{\ID}]-1 = \int_{\mathcal{S}_{\ID}} \!\!\!\! d\omega^{\xi}_{\mathcal{S}}f(\xi) - 1 ~,\\  
\mathfrak{C}_{jk} & = \mathbb{E}_{\mu_{ME}^{\ID}}[(Z^j)^{*}Z^k] - \rho_{jk}\nonumber\\
& = \int_{\mathcal{S}_{\ID}} \!\!\!\! d\omega^{\xi}_{\mathcal{S}}f(\xi)(Z^j)^{*}(\xi) Z^k(\xi) - \rho_{jk} 
  ~,
\end{align*}
\end{subequations}
where $Z(\xi) : \xi \in \mathcal{S}_\ID \to Z(\xi) \in \PHS$ are the homogeneous coordinate representation of the embedding 
functions $X^a$ of $\mathcal{S}_{\ID}$ onto $\PHS$.

The solution of the optimization problem leads to the Gaussian form, in homogeneous coordinates, with support on $\mathcal{S}_{\ID}$:
\begin{subequations}
\begin{align*}
&q_{ME}^{\ID}(\xi) = \frac{1}{Q_{\ID}}e^{-\sum_{j,k=1}^{d_S} \gamma_{jk}(Z^{j})^{*}(\xi)Z^k(\xi)}\\
&Q_\ID = \int_{\mathcal{S}_\ID}\!\!\!\! d\omega_{\mathcal{S}}^\xi~e^{-\sum_{j,k=1}^{d_S} \gamma_{jk}(Z^{j})^{*}(\xi)Z^k(\xi)}
  ~.
\end{align*}
\end{subequations}
Again, we can move from a homogeneous representation to a symplectic one $Z(\xi) \leftrightarrow \left(\vec{p}(\xi),\vec{\phi}(\xi)\right)$
in which the reference basis is the eigenbasis of $\rho$. This gives $\rho_{jk}
= \lambda_j \delta_{jk}$. This, in turn, means we only need
the diagonal Lagrange's multipliers $\gamma_{jj}$. As for the previous case, we
move to a single label notation $\gamma_{jj} \to \gamma_{j}$:
\begin{subequations}
\begin{align*}
&q_{ME}^{\ID}(\xi) = \frac{1}{Q_\ID(\vec{\tau})}e^{\vec{\tau} \cdot \vec{p}(\xi)}~,\\ 
& Q_\ID(\vec{\tau}) = \int_{\mathcal{S}_{\ID}}\!\!\!\!\! d\omega_{\mathcal{S}}^{\xi}~e^{\vec{\tau} \cdot \vec{p}(\xi)}\\
& \vec{\tau} \coloneqq \left( \gamma_{d_S}-\gamma_1,\gamma_{d_S}-\gamma_2,\ldots,\gamma_{d_S}-\gamma_{d_S-1}\right)
\end{align*}
\end{subequations}
with an analytical expression for the entropy:
\begin{align*}
\mathfrak{h}_{\ID}(\vec{\tau}) = \log Q_\ID(\vec{\tau}) - \vec{\tau} \cdot \vec{\lambda}
  ~.
\end{align*}

While this solution appears to have much in common with the $\ID = 2(d_S-1)$
case, there are profound differences. Indeed, the functions $\vec{p}(\xi)$ can
be highly degenerate, since we are embedding a low-dimensional manifold
$\mathcal{S}_{\ID}$ into a higher one $\PHS$. Indeed, the coordinates $\xi$
emerge from coordinatizing a submanifold of dimension $\ID$ within one of
dimension $2(d_S-1)$. This means that for $\mathcal{S}_{\ID}$ there are
$2(d_S-1)-\ID$ independent equations of the type
$\left\{K_n(Z)=0\right\}_{n=1}^{2(d_S-1)-\ID}$. In general, we expect them to
be highly nonlinear functions of their arguments. While choosing an appropriate
coordinate system allows simplifying, this choice has to be made on a
case-by-case basis. In specific cases, discussed in the next section, several
exact solutions can be found analytically.

\subsection*{Noninteger $\ID$: Fractal ensembles.}

As Ref. \cite{Anza22a} showed, even measuring the environment in a local
basis can lead to GQSs with noninteger $\ID$. For example, if we explicitly
break the translational invariance of the spin-1/2 Heisenberg model in $1D$ by
changing the local magnetic field of one spin, the GQS of one of its spin-$1/2$ is described by a fractal resembling Cantor's set in the thermodynamic limit of an infinite environment. Its quantum information dimension and
geometric entropy have been estimated numerically to be $\ID \approx 0.83 \pm
0.02$ and $\mathfrak{h}_{0.83}$ grows linearly with $N_E$, the size of the
environment: $\mathfrak{h}_{0.83} \propto 0.66 N_E$. Their existence gives
physical meaning to the question of finding the maximum geometric entropy
ensemble with noninteger $\ID$. 

Providing concrete solutions to this problem is quite complex, as it requires
having a generic parametrization for an ensemble with an arbitrary fractional
$\ID$. As far as we know, this is currently not possible. While we do know that
certain ensembles have a noninteger $\ID$, there is no guarantee that fixing
the value of the information dimension, e.g. $\ID = N/M$ with $N,M \in
\mathbb{N}$ relative primes, turns into an explicit way of parametrizing the
ensemble. We leave this problem open for future work.

\section{How does $\mu_{ME}$ emerge?}
\label{sec:Discussion}

While the previous section gave the technical details regarding ensembles
resulting from the proposed maximum geometric quantum entropy principle, the
following identifies the mechanisms for their emergence in a number of cases of
physical interest.

\subsection*{Emergence of $\mu_{ME}^{0}$.}  

As partly discussed in the previous section, $\mu_{ME}^{0}$ can emerge
naturally as a conditional ensemble, when our system of interest interacts with
a finite-dimensional environment (dimension $N$). If the environment is probed
with projective measurements in a basis that is unbiased with respect to the
Schmidt-partner basis $\left\{ \ket{SP_\alpha}\right\}_{\alpha=1}^{N}$, we
reach the absolute maximum of the geometric entropy, $\log N$. The resulting
GQS is $\mu_{ME}^{0} = \delta^{\vec{p}}_{\vec{\lambda}}
\frac{1}{N}\sum_{\alpha=1}^N \delta^{\vec{\phi}}_{\vec{\theta}_\alpha}$, with
members of the ensemble being $\ket{\chi_\alpha} = \sum_{j=1}^{d_S}
\sqrt{\lambda_j} \e^{i\theta_{\alpha j}}\ket{\lambda_j}$ and $p_\alpha = 1/N$.

As argued in Ref. \cite{Anza2018}, the notion of unbiasedness is typical.
Physically, this is interpreted as follows. Imagine someone gives up
$\ket{\psi(\rho)}$, a purification of $\rho$, without telling us anything about
the way the purification is done. This means we know nothing about the way
$\rho$ has been encoded into $\ket{\psi(\rho)}$. Equivalently, we do not know
what the $\left\{\ket{SP_j}\right\}_{j=1}^{d_S}$ are. If we now choose a basis
of the environment to study the conditional ensemble, $\left\{
\ket{v_\alpha}\right\}_{\alpha=1}^{N}$, this will have very little information
about the $\left\{\ket{SP_j}\right\}_{j=1}^{d_S}$---there is a very high chance
that we will end up very close to the unbiasedness condition.

The mathematically rigorous version of ``very high chance'' and ``very close''
is given in Ref. \cite{Anza2018} and it is not relevant here. The only thing we
need is that this behavior is usually exponential in the size of the
environment $\sim 2^{N}$. Somewhat more accurately, the fraction of bases which
are $\left| \braket{v_\alpha}{SP_j}\right|^2 \approx 1/N$ are $\sim 1-2^{-N}$.
Therefore, statistically speaking, it is extremely likely that, in absence of
meaningful information about what the $\left\{ \ket{SP_j}\right\}_{j=1}^{d_S}$
are, the conditional ensemble we will see is $\mu_{ME}^{0}$.

\subsection*{Emergence of $\mu_{ME}^{2(d_S-1)}$}

For $\mu_{ME}^{2(d_S-1)}$ to emerge as a conditional ensemble our
$d_S$-dimensional quantum system must interact with an environment that is
being probed with measurements whose outcomes are parametrized by $2(d_S-1)$
continuous variables, each with the cardinality of the reals. This is because
we have to guarantee that $\ID = 2(d_S-1)$. Therefore, conditioning on
projective measurements on a finite environment is insufficient. One
possibility is to have a finite environment that we measure on an overcomplete
basis, like coherent states. A second possibility is to have a genuinely
infinite-dimensional environment, on which we perform projective measurements.
For example, we could have $2(d_S-1)/3$ quantum particles in $3D$ that we
measure on the position basis
$\left\{\otimes_{n=1}^{2(d_S-1)/3}\ket{x_n,y_n,z_n}\right\}$. All the needed
details were given in Ref. \cite{Anza20a}, where we studied the properties of a
GQS emerging from a finite-dimensional quantum system interacting with one with
continuous variables.

We stress here that this is only a necessary condition, not a sufficient one.
Indeed, we can have an infinite environment that is probed with projective
measurements on variables with the right properties, but still obtain an
ensemble that is not $\mu_{ME}^{2(d_S-1)}$. An interesting example of this is
given by the continuous generalization of the notion of unbiased basis. We
illustrate this in a simple example of a purification obtained with a set of
$2(d_S-1)$ real continuous variables, realized by $2(d_S-1)$ non-interacting
particles in a $1D$ box $[0,L]$.

In this, the notion of an unbiased basis is satisfied by position and momentum
eigenstates: $\left\langle \vec{x} \vert \vec{k}\right\rangle = \frac{e^{i
\vec{k}\cdot \vec{x}}}{\sqrt{V}}$. Thus, if our Schmidt partners are momentum
eigenstates $\left\{\ket{SP_j} = \ket{\vec{k}_j}\right\}_{j=1}^{d_S}$, and we
measure the environment in the position basis, we do not obtain a GQS with the
required $\ID = 2(d_S-1)$. Indeed, while we do get $q(\vec{x}) =
\bra{\psi(\rho)} \mathbb{I}_S \otimes \ket{\vec{x}}\bra{\vec{x}}
\ket{\psi(\rho)} = \frac{1}{V}$, the members of the ensemble
$\ket{\chi(\vec{x})} = \sum_{j=1}^{d_S} \sqrt{\lambda_j}e^{i\vec{k}_j \cdot
\vec{x}}\ket{\lambda_j}$ are not distributed in the appropriate way.

This leads to the ensemble
$\delta^{\vec{p}}_{\vec{\lambda}}~\frac{1}{(2\pi)^{d_S-1}}$, which has the
wrong information dimension: $\ID = d_S-1$, not $\ID = 2(d_S-1)$. This
clarifies why, in order to have $\ID = 2(d_S-1)$, using an environmental basis
that is unbiased with respect to the Schmidt partners is not enough.
Specifically, the probabilities $p_j(\vec{x}) = \left\vert
\braket{\lambda_j}{\chi(\vec{x})} \right\vert^2 = \lambda_j$ do not depend on
$\vec{x}$. They do not get redistributed by the unbiasedness condition and are
always equal to the eigenvalues of $\rho$.

If we measure on a different basis $\left\{ \ket{\vec{l}} :=\int_V
d\vec{x}~u^{*}_{\vec{l}}(\vec{x})\right\}$, we obtain a different GQS since $\braket{\vec{l}}{SP_j} = \int_V d\vec{x}u_{\vec{l}}(\vec{x})e^{i\vec{k}_j \cdot \vec{x}} = \mathcal{F}_{\vec{l}}(\vec{k}_j)$ is essentially the Fourier transform of $u_{\vec{l}}$:
\begin{align}
&q(\vec{l}) = \sum_{j=1}^{d_S} \lambda_j \left\vert\mathcal{F}_{\vec{l}}(\vec{k}_j)\right\vert^2\\
& \ket{\chi(\vec{l})} = \sum_{j=1}^{d_S} \sqrt{\lambda_j} \frac{\mathcal{F}_{\vec{l}}(\vec{k}_j)}{\sqrt{q(\vec{l})}}\ket{\lambda_j}\label{eq:pphi}
  ~.
\end{align}
Equation (\ref{eq:pphi}) gives the functions $\left\{
p_j(\vec{l}),\phi_j(\vec{l})\right\}_{j=1}^{d_S-1}$:
\begin{align}
&p_j(\vec{l}) = \frac{\lambda_j \left\vert\mathcal{F}_{\vec{l}}(\vec{k}_j)\right\vert^2 }{\sum_{n=1}^{d_S} \lambda_n \left\vert\mathcal{F}_{\vec{l}}(\vec{k}_n)\right\vert^2}
  ~, \label{eq:pjl}\\
&\phi_j(\vec{l}) = \mathrm{Arg}\left( \mathcal{F}_{\vec{l}}(\vec{k}_j)\right)
  ~.\label{eq:phijl}
\end{align}
This, together with the density $q(\vec{l})$ specifies the ensemble via $\mu = \int d\vec{l} q(\vec{l}) \delta^{\vec{p}}_{\vec{p}(\vec{l})} \delta^{\vec{\phi}}_{\vec{\phi}(\vec{l})}$.

Finding the exact conditions that lead to $\mu = \mu_{ME}^{2(d_S-1)}$ involves
solving a complex inverse problem. However, what we have done so far allows us
to understand the real mechanism behind its emergence. First, the
$\phi_j(\vec{l})$ must be uniformly distributed: they must be random phases.
Second, the distribution of $\vec{p}$ must be of exponential form. The first
condition can always be ensured by choosing some $\braket{\vec{l}}{\vec{x}} =
u_{\vec{l}}(\vec{x})$ and then multiplying it by pseudo-random phases,
generated in a way that is completely independent on $\vec{p}$. This can always
be done without breaking the unitarity of $\braket{\vec{l}}{\vec{x}}$ via
$u_{\vec{l}}(\vec{x}) \to u_{\vec{l}}(\vec{x})e^{i\theta_{\vec{l}}}$. This
guarantees that the marginal distribution over the phases is uniform and that
the density $q(\vec{p},\vec{\phi})$ becomes a product of its marginals, since
the distribution of the $\vec{\phi}$ has been built to be independent of
everything else: $q(\vec{p},\vec{\phi}) =
f(\vec{p})~\cdot~\mathrm{unif}(\vec{\phi})$. Then, in order for
$q(\vec{p},\vec{\phi})$ to be the maximum entropy one we need $f(\vec{p}) =
\frac{1}{Q(\vec{\tau})}e^{\vec{\tau}\cdot \vec{p}}$.

Given a nondegenerate $\vec{p}(\vec{l})$, this can be ensured by a specific form of $q(\vec{l})$ since $f(\vec{p}) = \int d\vec{l} q(\vec{l})\delta^{\vec{p}}_{\vec{p}(\vec{l})}$:
\begin{align*}
q(\vec{l}) = \left(\mathrm{det} J \right)\!(\vec{l})~\frac{e^{\vec{\tau}\cdot \vec{p}(\vec{l})}}{Q(\tau)}~\Rightarrow~f(\vec{p}) = \frac{e^{\vec{\tau} \cdot \vec{p}}}{Q(\tau)}
  ~,
\end{align*}
where $J$ is the Jacobian matrix of the coordinate change $\vec{l} \to
\vec{p}(\vec{l})$. Checking that this form leads to the right distribution is
simply a matter of coordinate changes. Alternatively, it can be seen by
repeated use of the Laplace transform on the simplex, together with the result
$\frac{1}{Q(\vec{\tau})}\int_{\Delta_{d_S-1}}\!\!\! d \vec{p} e^{-\vec{a}\cdot
\vec{p}}e^{\vec{\tau}\cdot \vec{p}} = Q(\vec{\tau}-\vec{a})/Q(\vec{\tau})$.  We
now see the mechanism at play in a concrete way and how it leads to the maximum
entropy GQS $\mu_{ME}^{2(d_S-1)}$.

First, let's take the label $\vec{l} = (l_1,\ldots,l_{2(d_S-1)})$ and split it
in two $\vec{l} = (\vec{a},\vec{b})$ with $\vec{a} = (a_1,\ldots,a_{d_S-1})$
and $\vec{b} = (b_1,\ldots,b_{d_S-1})$. Then $d\vec{l} = d\vec{a} d\vec{b}$. At
this stage, the choice of $\vec{l}$, the splitting, and $\ket{SP_j}$ are
arbitrary. Then, we make the choice that $\braket{\vec{a},\vec{b}}{SP_j} =
\sqrt{A_j(\vec{a})}e^{iB_j(\vec{b})}$. The only property we need to check is
that $\left\{ \ket{\vec{a},\vec{b}}\right\}_{\vec{a},\vec{b}}$ can be a
complete set:
\begin{align*}
\int d\vec{a} \sqrt{A_j(\vec{a})A_k(\vec{a})} \int d\vec{b} e^{i(B_j(\vec{b})-B_k(\vec{b}))} = \delta_{jk}
  ~.
\end{align*}
We can choose $B_j(\vec{b})$ such that $\int d\vec{b} e^{i(B_j(\vec{b})-B_k(\vec{b}))} = M \delta_{jk}$, for example by choosing $B_j(\vec{b})$ 
to be linear functions. Then, choosing $A_j(\vec{a})$ such that $\int d\vec{a} A_j(\vec{a}) = \frac{1}{M}$ guarantees completeness. With this
choice, we obtain:
\begin{align*}
&q(\vec{a},\vec{b}) = \sum_{j=1}^{d_S} \lambda_j A_j(\vec{a})~,\\
&p_j(\vec{a},\vec{b}) = \frac{\lambda_j A_j(\vec{a})}{\sum_{n=1}^{d_S-1}\lambda_n A_n(\vec{a})} &&\to p_j(\vec{a})~,\\
&\phi_j(\vec{a},\vec{b}) = B_j(\vec{b}) &&\to \phi_j(\vec{b})
  ~.
\end{align*}

The probability density $q(\vec{a},\vec{b})$ can be written as a product of two probability densities: $q(\vec{a},\vec{b}) = f(\vec{a})\frac{1}{M}$.
Here, $1/M$ is the uniform density for $\vec{b}$ and $f(\vec{a}) = \sum_{j=1}^{d_S} \lambda_j M A_j(\vec{a})$ is a probability density for $\vec{a}$.
Then, the GQS becomes a product of two densities: one over the probability simplex (for $\vec{p}$) and another one over the phases (for $\vec{\phi}$):
\begin{align}
\mu &= \!\! \int \!\!d\vec{a}\!\!\int \!\! d \vec{b}~q(\vec{a},\vec{b})~\delta^{\vec{p}}_{\vec{p}(\vec{a},\vec{b})}\delta^{\vec{\phi}}_{\vec{\phi}(\vec{a},\vec{b})}~,\\
& = \int d\vec{a} f_a(\vec{a}) \delta^{\vec{p}}_{\vec{p}(\vec{a})}~\cdot~\frac{1}{M}\int d\vec{b} \delta^{\vec{\phi}}_{\vec{\phi}(\vec{b})}~,\\
& = f_p(\vec{p}) f_\phi(\vec{\phi})~.
\end{align}
These are, basically, the formulas for two changes of variable in integrals:
$\vec{a} \to \vec{p}(\vec{a})$ and $\vec{b} \to \vec{\phi}(\vec{b})$. Since
these are invertible, we can confirm what we understood before.
$f_{\phi}(\vec{\phi}) = \mathrm{unif}[\vec{\phi}]$ when the phases
$\vec{\phi}(\vec{b})$ are uniformly distributed. Moreover, when $f_p(\vec{p}) =
\frac{e^{\tau \cdot \vec{p}}}{Q(\vec{\tau})}$ and $\vec{p}(\vec{a})$ are
exponentially distributed: $f_a(\vec{a}) = \left(\mathrm{det}
J_{ap}\right)(\vec{a})\frac{e^{\vec{\tau}\cdot
\vec{p}(\vec{a})}}{Q(\vec{\tau})}$, with $J_{ap}$ being the Jacobian matrix of
the change of variables $\vec{a} \to \vec{p}(\vec{a})$.

\paragraph*{Stationary distribution of some dynamic.}
A second mechanism, that can lead to the emergence of an ensemble with $\ID =
2(d_S-1)$, is time averaging. Indeed, if we are in a nonequilibrium dynamical
situation in which the system and its environment jointly evolve with a
dynamical (possibly unitary) law, its conditional ensembles $\mu(t)$ depend on time.

To study stationary behavior from dynamics one looks at time-averaged
$\overline{\mu(t)} = \lim_{T \to \infty} \frac{1}{T}\int_0^T \mu(t)$ ensembles
that, in this case, have a certain stationary density matrix $\rho_{ss} =
\overline{\rho(t)}$. Unless something peculiar happens, we expect the ensemble
to cover large regions of the full state space, leading to a stationary GQS
with $\ID = 2(d_S-1)$ and a given density matrix $\rho_{ss}$.

Intuitively, we expect dynamics that are chaotic in quantum state space to lead
to ensembles described by $\mu_{ME}^{2(d_S-1)}$. This is because the ensemble
that emerges must be compatible with a density matrix $\rho_{ss}$, while still
exhibiting a nontrivial dynamics due to the action of the environment. We now
give a simple example of how this happens. Borrowing from Geometric Quantum
Thermodynamics, see Ref. \cite{Anza22} where we studied a qubit with a
Caldeira-Leggett-like environment. The resulting evolution for the qubit can be
described using Stochastic Schr\"odinger's equation which, as shown in Ref.
\cite{Anza22}, leads to a maximum entropy ensemble (see Eq.
(\ref{eq:me_max_D})) of the required type.

\subsection*{Emergence of $\mu_{ME}^{d_S-1}$}

Among all possible values of $\ID$, a third one which is particularly relevant
is $\ID = d_S-1$, which is half the maximum value. The reason why this is
important comes from the symplectic nature of the quantum state space and,
ultimately, from dynamics. One physical situation in which $\mu_{ME}^{d_S-1}$
emerges naturally is the study of the dynamics of pure, isolated quantum
systems. The phenomenology we discuss here is known, being intimately related
to thermalization and equilibration studies. We discuss it here only in
connection with the maximum geometric entropy principle introduced in
Section \ref{sec:max_geom_ent}.

Imagine an isolated quantum system in a pure state $\ket{\psi_0}$ evolving
unitarily with a dynamics generated by some time-independent Hamiltonian $H =
\sum_{n=1}^{D}E_n \ket{E_n}\bra{E_n}$. Assuming lack of degeneracies in the
energy spectrum, the dynamics is given by $\ket{\psi_t} =
\sum_{n=1}^{D}\sqrt{p_n^0}e^{i(\phi^0_n - E_n t)}$, where
$\sqrt{p^0_n}e^{i\phi^0_n}\coloneqq \braket{E_n}{\psi_0}$ and we have used
symplectic coordinates in the energy eigenbasis. Since $\vec{p} \in
\Delta_{D-1}$ are conserved quantities $p_n^t = p_n^0$ and $\vec{\phi} \in
\mathbb{T}_{D-1}$ evolve independently and linearly $\phi_n^t = \phi_n^0 - E_n
t$ on a high-dimensional torus, we know that a sufficient condition for the
emergence of ergodicity on $\mathbb{T}_{D-1}$ is the so-called non-resonance
condition: energy gaps have to be non-degenerate: namely $E_n - E_k = E_a -
E_b$ if and only if $n=k$ and $a=b$ or $n=a$ and $k=b$.

This condition is usually true for 
interacting many-body quantum systems. If that's the case, then the evolution of the phases is ergodic on $\mathbb{T}_{D-1}$.
This was first proven by von Neumann \cite{vNeu29} in 1929. Calling $(\vec{p}(t),\vec{\phi}(t))$ the instantaneous state 
and with $\delta^{\vec{p}}_{\vec{p}(t)} \delta^{\vec{\phi}}_{\vec{\phi}(t)}$
the corresponding Dirac measure on the quantum state space, we have:
\begin{align*}
\lim_{T \to \infty} \frac{1}{T}\int_0^T dt \delta^{\vec{p}}_{\vec{p}(t)} \delta^{\vec{\phi}}_{\vec{\phi}(t)} = \delta^{\vec{p}}_{\vec{p}(0)}~\cdot~\mathrm{unif}^{\vec{\phi}}\left(\mathbb{T}_{D-1}\right)
  ~,
\end{align*}
where $\mathrm{unif}^{\vec{\phi}}\left(\mathbb{T}_{D-1}\right)$ is the uniform
measure on $\mathbb{T}_{D-1}$ in which all $\phi_n$ are uniformly and
independently distributed on the circle. It is not too hard to see that this is
the maximum geometric entropy ensemble with $\ID = d_S-1$, compatible with the
fact that the occupations of the energy eigenstates are all conserved
quantities: $p_n^t = p_n^0$.

Indeed, these $d_S-1$ constraints provide $d_S-1$ independent equations, thus
reducing the state-space dimension that the system explores to the
high-dimensional torus $\mathbb{T}_{D-1}$. On this, however, the dynamics is
ergodic and the resulting stationary measure is the uniform one. By definition,
this is the measure with the highest possible value of geometric entropy since
its density is uniform and equal to $q_{ME}^{\phi}(\vec{\phi}) =
\frac{1}{\mathrm{Vol}[\mathbb{T}_{D-1}]}$, where
$\mathrm{Vol}[\mathbb{T}_{D-1}] = \int_{\mathbb{T}_{D-1}}\!\!\!
d\omega_{FS}^{\phi} = (2\pi)^{D-1}$ is the volume of $\mathbb{T}_{D-1}$,
computed with the Fubini-Study volume element projected on $\mathbb{T}_{D-1}$, that is
$\prod_{k=1}^{D-1} d\phi_k$:
\begin{align*}
\mathfrak{h}_{d_S-1} &= - \int_{\mathbb{T}_{D-1}}\!\!\!d\omega_{FS}^{\phi} \frac{1}{\mathrm{Vol}[\mathbb{T}_{D-1}]} \log \frac{1}{\mathrm{Vol}[\mathbb{T}_{D-1}]}\nonumber \\
& = \log \mathrm{Vol}[\mathbb{T}_{D-1}]
  ~.
\end{align*}

\subsection*{Comment on the generic $\mu_{ME}^{\ID}$.}

To have a GQS with generic information dimension $\ID$ result from a
conditional measurement on an environment, we must condition on at least $\ID$
continuous variables with the cardinality of the reals. This can be achieved
either via measurements on an overcomplete basis, such as coherent states, or
via projective measurements on a infinite dimensional environment with at least
$\ID$ real coordinates. This condition is necessary, but not sufficient, to
guarantee the emergence of the corresponding maximum entropy ensemble
$\mu_{ME}^{\ID}$. While we have seen that the notion of an unbiased basis is
relevant when $\ID = 0$, we also saw how this falls short in the generic $\ID >
0$ case. Understanding this general condition is a nontrivial task that
requires a much deeper understanding of how systems encode quantum information
in their environment and how this is extracted by means of quantum
measurements. Further work in this direction is ongoing and will be reported
elsewhere.

For a GQS with arbitrary dimension $\ID$ to emerge as a stationary distribution on a quantum state space with dimension
$2(d_S-1)$ it is likely that we need $2(d_S-1) - \ID$ independent equations constraining the dynamics. That is, if $\ID$ is an integer.
Indeed, due to the continuity of time and the smoothness of the time evolution in quantum state space, we expect $\ID \in \mathbb{N}$ 
in the vast majority of cases. If these equations constraining the motion on the quantum state space are linear, then we know that 
having $2(d_S-1) - \ID$ independent equations is both necessary and sufficient to have $\ID$ as quantum information dimension. 
This, however, says virtually nothing about the maximization of the relevant geometric entropy $\mathfrak{h}_{\ID}$. Moreover, constraints 
on an open quantum system can take very generic forms and the relevant equations will not always be linear. 

An explicit example where such ensemble can be found constructively is given by the case of an isolated quantum system. The conditions
for the emergence of a maximum entropy $\mu_{ME}^{d_S-1}$ are then known, being equivalent to the conditions for the ergodicity of
periodic dynamics on a high-dimensional torus, which are known.

\section{Conclusion}\label{sec:Conclusion}

While a density matrix  encodes all the statistics available from performing
measurements on a system, they do not give information about how the statistics
was created. And, infinite possibilities are available. A natural way to select
a unique ensemble behind a given density matrix is to approach the problem from
the perspective of information theory. In this case the issue becomes a
standard inference problem, one to which we can apply the maximum entropy
principle. To properly formulate the problem in this way requires a proper way
to compute the entropy of an ensemble. While this is trivial for ensembles with
a finite number of elements, it is not for continuous ensembles. The correct
answer, the notion of Geometric Quantum Entropy $\GE$, was given in
Ref.\cite{Anza22a}. This, however, depends strongly on another quantity that
characterizes the ensemble: the quantum information dimension $\ID$.
Consequently, we formulated the maximum geometric entropy principle at fixed
quantum information dimension. This is a one-parameter class of maximum entropy
principles, labeled by $\ID$, that can be used to explore various ways to have
ensembles give rise to a specific density matrix.

As often happens with inference principles, the generic optimization problem
can be hard to solve. However, here we solved a number of cases where the
ensemble can be found analytically. We also explored the physical mechanism
responsible for the emergence of $\mu_{ME}^{\ID}$. Two different classes of
situations were considered: (i) conditional ensemble, resulting from measuring
the environment of our system of interest, and (ii) stationary distributions,
in which the statistics arise from aggregating data over time. We have also
identified and discussed various instances where both mechanisms lead to a
maximum entropy ensemble.

\section*{Acknowledgments}
\label{sec:acknowledgments}

F.A. thanks Marina Radulaski, Davide Pastorello, Akram Touil, Sebastian Deffner
and Davide Girolami for discussions on geometric quantum mechanics. 
F.A. and J.P.C. thank David Gier and Ariadna Venegas-Li for helpful
discussions and the Telluride Science Research Center for its hospitality
during visits. This material is based upon work supported by, or in part by, a
Templeton World Charity Foundation Power of Information Fellowship, FQXi Grant
FQXi-RFP-IPW-1902, and U.S. Army Research Laboratory and the U. S. Army
Research Office under contracts W911NF-13-1-0390 and W911NF-18-1-0028.

\section*{Data Availability Statement}
The data that support the findings of this study are available from the
corresponding author upon reasonable request.

\bibliography{library}

\begin{thebibliography}{10}

\bibitem{Pathria2011}
R.~K. Pathria and Paul~D. Beale.
\newblock {\em {Statistical Mechanics}}.
\newblock Elsevier B.V., 2011.

\bibitem{Greiner1995}
W.~Greiner, L.~Neise, and H.~St{\"{o}}cker.
\newblock {\em {Thermodynamics and Statistical Mechanics}}.
\newblock Springer New York, New York, NY, 1995.

\bibitem{Jaynes1957}
E.~T. Jaynes.
\newblock {Information Theory and Statistical Mechanics}.
\newblock {\em Phys. Rev.}, 106(4), 1957.

\bibitem{Jaynes1957a}
E.~T. Jaynes.
\newblock {Information Theory and Statistical Mechanics. II}.
\newblock {\em Phys. Rev.}, 108:171--190, 1957.

\bibitem{Cover2006}
T.~M. Cover and J.~A. Thomas.
\newblock {\em {Elements of Information theory}}.
\newblock Wiley-Interscience, New York, 2006.

\bibitem{Anza20a}
F.~Anza and J.~P. Crutchfield.
\newblock Beyond density matrices: Geometric quantum states.
\newblock {\em Phys. Rev. A}, 2020.

\bibitem{Anza22}
F.~Anza and J.~P. Crutchfield.
\newblock Geometric quantum thermodynamics.
\newblock {\em Phys. Rev. E}, 2022.

\bibitem{STROCCHI1966}
F.~Strocchi.
\newblock {Complex Coordinates and Quantum Mechanics}.
\newblock {\em Rev. Mod. Physics}, 38(1):36--40, 1966.

\bibitem{Kibble1979}
T.~W.~B. Kibble.
\newblock {Geometrization of quantum mechanics}.
\newblock {\em Comm. Math. Physics}, 65(2):189--201, 1979.

\bibitem{Heslot1985}
A.~Heslot.
\newblock {Quantum mechanics as a classical theory}.
\newblock {\em Phys. Rev. D}, 31(6):1341--1348, 1985.

\bibitem{Gibbons1992}
G.~W. Gibbons.
\newblock {Typical states and density matrices}.
\newblock {\em J. Geom. Physics}, 8(1-4):147--162, 1992.

\bibitem{Ashtekar1995}
A.~Ashtekar and T.~A. Schilling.
\newblock {Geometry of quantum mechanics}.
\newblock In {\em AIP Conference Proceedings}, volume 342, pages 471--478. AIP,
  1995.

\bibitem{Ashtekar1999}
A.~Ashtekar and T.~A. Schilling.
\newblock {Geometrical formulation of quantum mechanics}.
\newblock In {\em On Einstein's Path}, pages 23--65. Springer New York, New
  York, NY, 1999.

\bibitem{Brody2001}
D.~C. Brody and L.~P. Hughston.
\newblock {Geometric quantum mechanics}.
\newblock {\em J. Geom. Physics}, 38(1):19--53, 2001.

\bibitem{Bengtsson2017}
I.~Bengtsson and K.~Zyczkowski.
\newblock {\em {Geometry of Quantum States}}.
\newblock Cambridge University Press, Cambridge, 2017.

\bibitem{Carinena2007}
J.~F. Cari{\~{n}}ena, J.~Clemente-Gallardo, and G.~Marmo.
\newblock {Geometrization of quantum mechanics}.
\newblock {\em Theoret. Math. Physics}, 152(1):894--903, 2007.

\bibitem{Chruscinski2006}
D.~Chru{\'{s}}ci{\'{n}}ski.
\newblock {Geometric aspects of quantum mechanics and quantum entanglement}.
\newblock {\em J. Physics Conf. Series}, 30:9--16, 2006.

\bibitem{Marmo2010}
G~Marmo and G~F Volkert.
\newblock {Geometrical description of quantum mechanics—transformations and
  dynamics}.
\newblock {\em Physica Scripta}, 82(3):038117, 2010.

\bibitem{Avron2020}
J.~Avron and O.~Kenneth.
\newblock {An elementary introduction to the geometry of quantum states with
  pictures}.
\newblock {\em Rev. Math. Physics}, 32(02):2030001, 2020.

\bibitem{Pastorello2015}
D.~Pastorello.
\newblock {A geometric Hamiltonian description of composite quantum systems and
  quantum entanglement}.
\newblock {\em Intl. J. Geom. Meth. Mod. Physics}, 12(07):1550069, 2015.

\bibitem{Pastorello2015a}
D.~Pastorello.
\newblock {Geometric Hamiltonian formulation of quantum mechanics in complex
  projective spaces}.
\newblock {\em Intl. J. Geom. Meth. Mod. Physics}, 12(08):1560015, 2015.

\bibitem{Pastorello2016}
D.~Pastorello.
\newblock {Geometric Hamiltonian quantum mechanics and applications}.
\newblock {\em International Journal of Geometric Methods in Modern Physics},
  13(Supp. 1):1630017, 2016.

\bibitem{Clemente-Gallardo2013}
J.~Clemente-Gallardo and G.~Marmo.
\newblock {The Ehrenfest picture and the geometry of quantum mechanics}.
\newblock {\em Il Nuovo Cimento C}, 3:35--52, 2013.

\bibitem{HJW93}
Lane~P. Hughston, Richard Jozsa, and William~K. Wootters.
\newblock A complete classification of quantum ensembles having a given density
  matrix.
\newblock {\em Physics Letters A}, 183(1):14--18, 1993.

\bibitem{Wise01}
J.~Vaccaro H.~Wiseman.
\newblock Inequivalence of pure state ensembles for open quantum systems: the
  preferred ensembles are those that are physically realizable.
\newblock {\em Phys. Rev. Lett.}, 2001.

\bibitem{Gold06}
S.~Goldstein, J.~L. Lebowitz, R.~Tumulka, and N.~Zanghi.
\newblock On the distribution of the wave function for systems in thermal
  equilibrium.
\newblock {\em Journal of Statistical Physics}, 2006.

\bibitem{Brody2000}
D.~C. Brody and L.~P. Hughston.
\newblock {Information content for quantum states}.
\newblock {\em J. Math. Physics}, 41(5):2586--2592, 2000.

\bibitem{Kar11}
R.~I. Karasik and H.~M. Wiseman.
\newblock How many bits does it take to track an open quantum system?
\newblock {\em Phys. Rev. Lett.}, 106:020406, Jan 2011.

\bibitem{Schro27}
Schroedinger.
\newblock The exhcange of energy in wave mechanics.
\newblock {\em Annalen der Physik}, 1927.

\bibitem{Schrodinger1989}
Erwin Schr{\"{o}}dinger.
\newblock {\em {Statistical thermodynamics}}.
\newblock Dover Publications, 1989.

\bibitem{Wal89}
Walecka.
\newblock {\em {Fundamentals of Statistical Mechanics. Manuscript and notes by
  Felix Bloch}}.
\newblock Stanford University Press, 1989.

\bibitem{Gold16}
Sheldon Goldstein, Joel~L. Lebowitz, Christian Mastrodonato, Roderich Tumulka,
  and Nino Zanghi.
\newblock Universal probability distribution for the wave function of a quantum
  system entangled with its environment.
\newblock {\em Communications in Mathematical Physics}, 342(3):965--988, March
  2016.

\bibitem{Reim08}
Peter Reimann.
\newblock Typicality of pure states randomly sampled according to the gaussian
  adjusted projected measure.
\newblock {\em Journal of Statistical Physics}, 132(5):921--935, September
  2008.

\bibitem{Anza22a}
F.~Anza and J.~P. Crutchfield.
\newblock Quantum information dimension and geometric entropy.
\newblock {\em Phys. Rev. X Quatum}, 2022.

\bibitem{Pollard02}
D.~Pollard.
\newblock {\em A user's guide to measure theoretic probability}.
\newblock Cambridge University Press, 2002.

\bibitem{Kolmogorov1933}
A.~Kolmogorov.
\newblock {\em Foundations of the theory of probability}.
\newblock Chelsea Publishing Company, 1933.

\bibitem{Vourdas04}
Vourdas.
\newblock Quantum systems with finite hilbert spaces.
\newblock {\em Rep. Prog. Phys.}, 67, 2004.

\bibitem{Bengtsson07}
Bengtsson.
\newblock Three ways to look at mutually unbiased bases.
\newblock {\em AIP Conf. Proc.}, 889, 2007.

\bibitem{Lawrence02}
Zeilinger Lawrence, Brukner.
\newblock Mutually unbiased binary observable sets on n qubits.
\newblock {\em Phys. Rev. A}, 65:032320, Feb 2002.

\bibitem{GordRodr20}
Elliott Gordon-Rodriguez, Gabriel Loaiza-Ganem, and John Cunningham.
\newblock The continuous categorical: a novel simplex-valued exponential
  family.
\newblock In Hal~Daumé III and Aarti Singh, editors, {\em Proceedings of the
  37th International Conference on Machine Learning}, volume 119 of {\em
  Proceedings of Machine Learning Research}, pages 3637--3647. PMLR, 2020.

\bibitem{Anza2018}
Fabio Anza, Christian Gogolin, and Marcus Huber.
\newblock {Eigenstate Thermalization for Degenerate Observables}.
\newblock {\em Physical Review Letters}, 120(15):150603, apr 2018.

\bibitem{vNeu29}
J.~v. Neumann.
\newblock Beweis des {Ergodensatzes} und {desH}-{Theorems} in der neuen
  {Mechanik}.
\newblock {\em Zeitschrift für Physik}, 57(1):30--70, January 1929.

\end{thebibliography}

\makeatletter
\newcommand{\manuallabel}[2]{\def\@currentlabel{#2}\label{#1}}
\makeatother

\manuallabel{sm:ETHs}{Section~A}
\manuallabel{sm:proof}{Section~B}
\manuallabel{sm:generalizedbloch}{Subsection~B1}
\manuallabel{sm:proofofmaintheorem}{Subsection~B2}
\manuallabel{sm:examples}{Section~C}
%\end{document}

\clearpage
\appendix
\onecolumngrid

\pagestyle{empty}

\begin{center}
{\Large {\bf \ourTitle}\\[0.5cm]
Supplemental Material}\\[0.2cm]
Fabio Anza and James P. Crutchfield
\end{center}

%\manuallabel{sm:PartitionFunction}{Appendix~A}
\section{Calculating the Partition Function}
\label{App:DetailedCalculation}

Recall:
\begin{align*}
\mathcal{Z}\left(\lambda_{\alpha\beta}\right)
  & = \int_{\mathcal{P}(\mathcal{H})}
  e^{- \sum_{\alpha,\beta} \lambda_{\alpha\beta}Z^\alpha \overline{Z}^\beta} dV_{FS} 
  ~.
\end{align*}
Since $\lambda_{\alpha\beta}$ are the Lagrange multipliers of
$\mathcal{C}_{\alpha\beta}$ we chose them to be Hermitian as they are not all
independent. Thus, we can always diagonalize them with a unitary matrix:
\begin{align*}
\sum_{\alpha \beta} U_{\gamma \alpha}\lambda_{\alpha\beta}U^{\dagger}_{\beta \epsilon} = l_\gamma \delta_{\gamma \epsilon}
  ~.
\end{align*}
This allows us to define auxiliary integration variables $X_\gamma =
\sum_{\alpha} U_{\gamma \alpha}Z^\alpha$. Thanks to these, we express the
quadratic form in the exponent of the integrand using that $(U^\dagger
U)_{\alpha \beta} = \delta_{\alpha \beta}$:
\begin{align*}
Z \lambda \overline{Z}
  & = \sum_{\alpha \beta}Z^\alpha \lambda_{\alpha \beta}\overline{Z}^\beta \\
  & = \sum_{\alpha \beta} \sum_{\tilde{\alpha} \tilde{\beta}}Z^\alpha
  \delta_{\alpha \tilde{\alpha}} \lambda_{\tilde{\alpha}
  \tilde{\beta}}\delta_{\beta \tilde{\beta}}\overline{Z}^{\tilde{\beta}} \\
  & = \sum_{\alpha \beta} \sum_{\tilde{\alpha} \tilde{\beta}} \sum_{ab}
  \left(Z^\alpha U^\dagger_{\alpha a}\right) \left( U_{a\tilde{\alpha}}
  \lambda_{\tilde{\alpha}\tilde{\beta}} U^\dagger_{\tilde{\beta}b}\right)
  \left( U_{b\beta}\overline{Z}^\beta\right) \\
  & = \sum_a \left\vert X_a \right\vert^2 l_a
  ~.
\end{align*}
Moreover, recalling that the Fubini-Study volume element is invariant under
unitary transformations, we can simply adapt our coordinate systems to $X_a$.
And so, we have $X_a = q_a e^{i\nu_a}$. This gives $dV_{FS} =
\prod_{k=1}^{D-1}\frac{dq_a d\nu_a}{2}$. We get to the following simpler
functional:
\begin{align*}
\mathcal{Z}\left(\lambda_{\alpha\beta}\right)
  & = \int_{\mathcal{P}(\mathcal{H})} e^{- \sum_{a=0}^{D-1} l_a q_a}
  \prod_{k=1}^{D-1}\frac{dq_a d\nu_a}{2} \\
  & = \frac{(2\pi)^{D-1}}{2^{D-1}} \int_{\Delta_{D-1}} e^{-\sum_a l_a q_a} \prod_a dq_a
  ~.
\end{align*}
Now, we are left with an integral of an exponential function over the $D-1$-simplex. We can use Laplace transform trick to solve this kind of integral:
\begin{align*}
I_{D-1}(r)
  & \coloneqq \int_{\Delta_{D-1}} \prod_{k=0}^{D-1} e^{- l_k q_k} \delta
  \left(\sum_{k=0}^{D-1} q_k-r\right) dq_1 \ldots dq_{D-1} \\
  \Rightarrow 
  \tilde{I}_{D-1}(z) & \coloneqq \int_0^{\infty} e^{-zr}I_{D-1}(r) dr ~,\\
\tilde{I}_n(z) & = \prod_{k=0}^n \frac{(-1)^k}{( l_k + z)} \\
  & = (-1)^{\frac{n(n+1)}{2}} \prod_{k=0}^n \frac{1}{z-z_k}
  ~,
\end{align*}
with $z_k = - l_k \in \mathbb{R}$.

The function $\tilde{I}_n(z)$ has $n+1$ real and distinct poles: $z=z_k =
-l_k$. Hence, we exploit the partial fraction decomposition of
$\tilde{I}_n(z)$, which is:
\begin{align*}
(-1)^{\frac{n(n+1)}{2}} \prod_{k=0}^n \frac{1}{z-z_k}
  = (-1)^{\frac{n(n+1)}{2}} \sum_{k=0}^n \frac{R_k}{z-z_k}
  ~,
\end{align*}
where:
\begin{align*}
R_k & = \left[ (z-z_k)\tilde{I}_n(z)\right]_{z=z_k} \\
    & = \prod_{j=0, \, j \neq k}^n\frac{(-1)^{\frac{n(n+1)}{2}}}{z_k-z_j}
  ~.
\end{align*}

Exploiting linearity of the inverse Laplace transform plus the basic result:
\begin{align*}
\mathcal{L}^{-1}\left[\frac{1}{s+a}\right](t) = e^{-at}\Theta(t)
  ~,
\end{align*}
where:
\begin{align*}
\Theta(t) & = \begin{cases}
  1 & t \geq 0 \\
  0 & t < 0
\end{cases}
  ~.
\end{align*}
We have for:
\begin{align*}
I_n(r) & = \mathcal{L}^{-1}[\tilde{I}_n(z)](r) \\
  & = \Theta(r) \sum_{k=0}^n R_k e^{z_k r}
  ~.
\end{align*}
And so:
\begin{align*}
Z & = I_{D-1}(1) \\
  & = \sum_{k=0}^{D-1}
  \frac{ e^{-l_k} }{\prod_{j=0, \,\, j \neq k}^{D-1} (l_k - l_j)}
  ~.
\end{align*}
Now, consider that $l_a$ are linear functions of the true matrix elements:
\begin{align*}
l_a & = f_a(\lambda_{\alpha \beta}) \\
  & = \sum_{\alpha \beta}U_{a\alpha} \lambda_{\alpha \beta} U^\dagger_{\beta a}
  ~.
\end{align*}
We arrive at:
\begin{align*}
\mathcal{Z}(\lambda_{\alpha \beta}) = \frac{ e^{-f_k(\lambda_{\alpha\beta})} }{\prod_{j=0, \,\, j \neq k}^{D-1} (l_k(\lambda_{\alpha\beta}) - l_j(\lambda_{\alpha\beta}))}
  ~.
\end{align*}

\section{Calculating Lagrange Multipliers}
\label{App:Tau}

Given the expression of the partition function, we now show that that the value
of Lagrange's multipliers $\gamma_{jk}$ can be given analytically, by extending
the Laplace transform technique exploited Appendix
\ref{App:DetailedCalculation}.

The nonlinear equation to fix $\gamma_{jk}$ is:
\begin{align*}
\frac{\partial \log Q}{\partial \gamma_{jk}} = -\frac{1}{Q}\int_{\mathcal{P}(\mathcal{H}_S)}\!\!\!\!\!\! dV_{FS} (Z^{j})^{*}Z^k e^{-\sum_{a,b}\gamma_{ab}(Z^{a})^{*}Z^b} = \rho_{jk}
  ~.
\end{align*}
We now use as reference basis the eigenbasis of $\rho$ and as coordinate system $(\vec{p},\vec{\phi})$. This means only the diagonals $\gamma_{jj}$ enter the equation.
\begin{align*}
&-\frac{\partial \log Q}{\partial \gamma_{kk}} = \lambda_k \to  - \frac{1}{Q}\int_{\Delta_{d_S-1}} \prod_{n=1}^{d_S-1}dp_n~p_k~e^{-\sum_{a}\gamma_{aa}p_a} = - \frac{1}{Q}\int_{\mathbb{R}^{d_S}_{+}} \prod_{n=1}^{d_S}dp_n~p_k~e^{-\sum_{a=1}^{d_S}\gamma_{aa}p_a} \delta\left(\sum_j p_j - 1\right)
  ~.
\end{align*}
To compute this we use the same Laplace transform technique we used before,
with a minor adaptation. First we do single lable notation $\gamma_{jj} \to
l_j$, then we define:
\begin{align*}
J_{D-1}^{(k)}(r)\coloneqq -\int_{\mathbb{R}^D_{+}} \left( \prod_{n=1}^D dp_n\right)~p_k~\left(\prod_{j=1}^D e^{-l_j p_j}\right) \delta \left( \sum_{k=1}^D p_k - r\right)
\end{align*}
Its Laplace transform is;
\begin{align*}
\tilde{J}_{D-1}^{(k)}(z) = & \int_0^\infty e^{-zr}J_{D-1}^{(k)}(r) = -\int_{\mathbb{R}^D_+} \left( \prod_{n=1}^D dp_n\right)~p_k~\left(\prod_{j=1}^D e^{-(l_j+z) p_j}\right) \\
= & \prod_{n \neq k}\left( \int_{0}^{\infty} \!\!\!dp_n e^{-(l_n+z)p_n}\right) \times \left( \int_0^{\infty} \!\!\! dp_k ~(-p_k) e^{-(l_k+z)p_k}\right)\\
= & \prod_{n=1}^{D} \left( \int_{0}^{\infty} \!\!\!dp_n e^{-(l_n+z)p_n}\right) \times \frac{\partial}{\partial l_k} \log \left( \int_{0}^{\infty} \!\!\!dp_k e^{-(l_k+z)p_k}\right)\\
= & \left(\prod_{n=1}^D G_n(z)\right) \frac{\partial \log G_k(z)}{\partial l_k}
\end{align*}
Where:
\begin{align*}
&G_k(z)\coloneqq \int_{0}^{\infty} \!\!\!dp_k e^{-(l_k+z)p_k} = \frac{1}{l_k+z}\int_0^{\infty} dy e^{-y} = \frac{1}{l_k+z} && \frac{\partial \log  G_k(z)}{\partial l_k} = - G_k(z)
\end{align*}
Therefore, we obtain:
\begin{align*}
& \tilde{J}_{D-1}^{(k)}(z) = -\left(\prod_{n\neq k}^{D}G_n(z)\right) \times G_k(z)^2 = - \left(\prod_{n \neq k} \frac{1}{z-z_n}\right) \frac{1}{(z-z_k)^2} \qquad \mathrm{where} \qquad z_n = -l_n
  ~.
\end{align*}
This can be written again as a sum, using the partial fraction decomposition:
\begin{align*}
& \tilde{J}_{D-1}^{(k)}(z) = - \left(\prod_{n \neq k} \frac{1}{z-z_n}\right) \frac{1}{(z-z_k)^2} = \sum_{n\neq k}\frac{R_n}{z-z_n} + \frac{R_k^{(1)}}{(z-z_k)^2}~,
\end{align*}
where:
\begin{align*}
&R_n = \left[ (z-z_n)\tilde{J}_{D-1}^{(k)}(z)\right]_{z=z_n} = \left(\prod_{j\neq n,k} \frac{1}{z_n - z_j}\right)\frac{1}{(z_n-z_k)^2} = \left(\prod_{j\neq n} \frac{1}{z_n - z_j}\right)\frac{1}{(z_n-z_k)}\\
&R_k^{(1)} = \left[ (z-z_k)^2\tilde{J}_{D-1}^{(k)}(z)\right]_{z=z_k} = \prod_{j\neq k} \frac{1}{z_k-z_j}
  ~.
\end{align*}
Exploiting the basic Laplace transform result:
\begin{align*}
\mathcal{L}^{-1}\left[\frac{1}{\left(s+a\right)^n}\right](t) = \frac{t^{n-1}e^{-at}}{\Gamma(n)}\Theta(t)
  ~,
\end{align*}
we can the invert the relation to compute $J_{D-1}(r=1)$:
\begin{align*}
J_{D-1}(r) = \mathcal{L}^{-1}\left[ \tilde{J}_{D-1}^{(k)}(z)\right](r) = \sum_{j\neq k} R_j e^{z_jr} + rR_k^{(1)}e^{z_kr}
  ~.
\end{align*}
Eventually, remembering that $z_k = - l_{k} = - \gamma_{kk}$, we obtain:
\begin{align*}
J_{D-1}^{(k)}(1) = \sum_{j\neq k} R_j e^{-\gamma_{jj}} + R_k^{(1)}e^{-\gamma_{kk}} = (-1)^{D}\sum_{j \neq k} \frac{e^{-\gamma_{jj}}}{\left[\prod_{a \neq j}(\gamma_{jj}-\gamma_{aa})\right] (\gamma_{jj}-\gamma_{kk})} - \frac{e^{-\gamma_{kk}}}{\prod_{a\neq k}(\gamma_{kk}-\gamma_{aa})}
  ~.
\end{align*}
The Lagrange's multipliers $\gamma_j$ can then be fixed by solving:
\begin{align*}
J_{D-1}^{(k)}(1) = - \lambda_k
  ~,
\end{align*}
where $\lambda_k$ are the eigenvalues of the density matrix: $\rho_{jk} = \delta_{jk} \lambda_k$.

\end{document}